\documentclass[aps,prd,twocolumn,showkeys,amsmath,amssymb]{revtex4}
\usepackage{graphicx}
\usepackage{epstopdf}
\usepackage{multirow}
\usepackage{subfigure}
\usepackage[colorlinks,citecolor=blue,anchorcolor=red,menucolor=red,linkcolor=red,filecolor=red,runcolor=red,urlcolor=blue,frenchlinks=red]{hyperref}

\allowdisplaybreaks[3]

\begin{document}

\title{Decay properties of the $Z_c(3900)$ through the Fierz rearrangement}

\author{Hua-Xing Chen}
\email{hxchen@buaa.edu.cn}
\affiliation{
School of Physics, Beihang University, Beijing 100191, China \\
School of Physics, Southeast University, Nanjing 210094, China
}

\begin{abstract}
We systematically construct all the tetraquark currents/operators of $J^{PC} = 1^{+-}$ with the quark configurations $[cq][\bar c \bar q]$, $[\bar c q][\bar q c]$, and $[\bar c c][\bar q q]$ ($q=u/d$), and derive their relations through the Fierz rearrangement of the Dirac and color indices. Using the transformations of $[qc][\bar q \bar c] \to [\bar c c][\bar q q]$ and $[\bar c q][\bar q c]$, we study decay properties of the $Z_c(3900)$ as a compact tetraquark state; while using the transformation of $[\bar c q][\bar q c] \to [\bar c c][\bar q q]$, we study its decay properties as a hadronic molecular state.
\end{abstract}

\maketitle

\section{Introduction}
\label{sec:intro}

In the past twenty years many charmonium-like $XYZ$ states were discovered in particle experiments~\cite{pdg}. All of them are good multiquark candidates, and their relevant experimental and theoretical studies have significantly improved our understanding of the strong interaction at the low energy region. Especially, in 2013 BESIII reported the $Z_c(3900)^+$ in the $Y(4260) \to J/\psi\pi^+\pi^-$ process~\cite{Ablikim:2013mio}, which was later confirmed by Belle~\cite{Liu:2013dau} and CLEO~\cite{Xiao:2013iha}. Since it couples strongly to the charmonium and yet it is charged, the $Z_c(3900)^+$ is not a conventional charmonium state and contains at least four quarks. It is quite interesting to understand how it is composed of these four quarks, and there have been various models developed to explain this, such as a compact tetraquark state composed of a diquark and an antidiquark~\cite{Maiani:2004vq,Maiani:2014aja}, a loosely-bound hadronic molecular state composed of two charmed mesons~\cite{Voloshin:1976ap,Voloshin:2013dpa,Wang:2013cya,Guo:2013sya,Wilbring:2013cha,He:2014nya,Chen:2015igx}, a hadro-quarkonium~\cite{Voloshin:2013dpa,Braaten:2013boa,Dubynskiy:2008mq}, or due to the kinematical threshold effect~\cite{Chen:2011xk,Chen:2013coa,Liu:2013vfa,Swanson:2014tra}, etc. We refer to reviews~\cite{Liu:2019zoy,Karliner:2017qhf,Guo:2017jvc,Olsen:2017bmm,Brambilla:2019esw} for detailed discussions.

The charged charmonium-like state $Z_c(3900)$ of $J^{PC}=1^{+-}$~\cite{Collaboration:2017njt} has been observed in the $J/\psi \pi$ and $D \bar D^*$ channels~\cite{Ablikim:2013mio,Liu:2013dau,Ablikim:2013xfr,Ablikim:2015swa}, and there was some events in the $h_c \pi$ channel~\cite{Ablikim:2013wzq}. In a recent BESIII experiment~\cite{Ablikim:2019ipd}, evidence for the $Z_c(3900) \rightarrow \eta_c\rho$ decay was reported with a statistical significance of $3.9\sigma$ at $\sqrt{s} = 4.226$ GeV, and the relative branching ratio
\begin{equation}
{\mathcal R}_{Z_c} \equiv {\mathcal{B}(Z_c(3900) \rightarrow \eta_c\rho) \over \mathcal{B}(Z_c(3900) \rightarrow J/\psi\pi)} \, ,
\end{equation}
was evaluated to be $2.2 \pm 0.9$ at the same center-of-mass energy. This ratio has been studied by many theoretical methods/models~\cite{Dias:2013xfa,Agaev:2016dev,Wang:2017lot,Goerke:2016hxf,Ke:2013gia,Patel:2014zja,Li:2014pfa,Dong:2013iqa,Faccini:2013lda,Xiao:2018kfx}, and was suggested in Ref.~\cite{Esposito:2014hsa} to be useful to discriminate between the compact tetraquark and hadronic molecule scenarios. As summarized in Table~\ref{tab:ratio}, this ratio was calculated in many molecular models, but the extracted values are highly model dependent. Hence, it would be useful to derive a model independent result, and it would be even better if one could do this within the same framework for both the tetraquark and molecule scenarios.

\begin{table*}[hbt]
\begin{center}
\renewcommand{\arraystretch}{1.5}
\caption{The relative branching ratio ${\mathcal R}_{Z_c} \equiv \mathcal{B}(Z_c(3900) \rightarrow \eta_c\rho) / \mathcal{B}(Z_c(3900) \rightarrow J/\psi\pi)$, calculated by various theoretical methods/models.}
\begin{tabular}{ c | c | c}
\hline\hline
~~~Interpretations~~~ & ~~~~~~~~~~~${\mathcal R}_{Z_c}$~~~~~~~~~~~ & ~~~Methods/Models~~~
\\ \hline\hline
                            &  $\left( 2.3^{+3.3}_{-1.4} \right) \times 10^2$      &  Type-I diquark-antidiquark model~\mbox{\cite{Esposito:2014hsa}}
\\ \cline{2-3}
\multirow{2}{*}{compact}
                            &  $0.27^{+0.40}_{-0.17}$                              &  Type-II diquark-antidiquark model~\mbox{\cite{Esposito:2014hsa}}
\\ \cline{2-3}
\multirow{2}{*}{tetraquark}
                            &  $0.95$                                              &  QCD sum rules~\mbox{\cite{Dias:2013xfa}}
\\ \cline{2-3}
                            &  $0.57$                                              &  QCD sum rules~\mbox{\cite{Agaev:2016dev}}
\\ \cline{2-3}
                            &  $1.1$                                              &  QCD sum rules~\mbox{\cite{Wang:2017lot}}
\\ \cline{2-3}
                            &  $1.28$                                              &  covariant quark model~\mbox{\cite{Goerke:2016hxf}}
\\ \hline\hline
                            &  $\left( 4.6^{+2.5}_{-1.7} \right) \times 10^{-2}$   &  Non-Relativistic effective field theory~\mbox{\cite{Esposito:2014hsa}}
\\ \cline{2-3}
{hadronic}
                            &  $0.12$                                              &  light front model~\mbox{\cite{Ke:2013gia}}
\\ \cline{2-3}
{molecule}
                            &  $0.68 \times 10^{-2}$                               &  effective field theory~\mbox{\cite{Patel:2014zja}}
\\ \cline{2-3}
                            &  $1.78$                                              &  covariant quark model~\mbox{\cite{Goerke:2016hxf}}
\\ \hline\hline
\end{tabular}
\label{tab:ratio}
\end{center}
\end{table*}

In this paper we shall study decay properties of the $Z_c(3900)$ under both the compact tetraquark and hadronic molecule interpretations. The present study is based on our previous finding that the diquark-antidiquark currents ($[qq][\bar q \bar q]$) and the meson-meson currents ($[\bar q q][\bar q q]$) are related to each other through the Fierz rearrangement of the Dirac and color indices~\cite{Chen:2006hy,Chen:2006zh,Chen:2007xr,Chen:2008ej,Chen:2008qw,Chen:2009tpa,Jiao:2009ra,Chen:2012ut,Chen:2013gnu,Chen:2013jra,Cui:2019roq}. More studies on light baryon operators can be found in Refs.~\cite{Chen:2008qv,Chen:2009sf,Chen:2012ex}. In the present case the $Z_c(3900)$ contains four quarks, that is the $c$, $\bar c$, $q$, $\bar q$ quarks ($q=u/d$), so there are three configurations:
\begin{equation*}
[cq][\bar c \bar q]\, , ~~ [\bar c q][\bar q c]\, , ~~ {\rm and} ~~ [\bar c c][\bar q q] \, .
\end{equation*}
Again, the Fierz rearrangement can be applied to relate them. Based on these relations, we shall extract some decay properties of the $Z_c(3900)$ in this paper.

There are eight independent $[cq][\bar c \bar q]$ currents of $J^{PC}=1^{+-}$, which have been systematically constructed in Ref.~\cite{Chen:2010ze}. Here we choose one of them,
\begin{equation}
\eta^{\mathcal Z}_\mu = \epsilon^{abe} \epsilon^{cde}~q_a^T\mathbb{C}\gamma_\mu c_b ~ \bar q_c \gamma_5\mathbb{C}\bar c_d^T - \{ \gamma_\mu \leftrightarrow \gamma_5 \} \, ,
\label{eq:j}
\end{equation}
where $\mathbb{C}$ is the charge-conjugation matrix, the subscripts $a \cdots e$ are color indices, and the sum over repeated indices is taken.
This current would strongly couple to the $Z_c(3900)$, if it has the same internal structure (internal symmetry) as that state.

The above current is useful from the viewpoints of both effective field theory and QCD sum rules. Note that there are various quark-based effective field theories, which have been successfully applied to describe the meson and baryon systems, such as the Non-Relativistic QCD for the heavy quarkonium system~\cite{Caswell:1985ui,Bodwin:1994jh}:
\begin{eqnarray}
\nonumber \mathcal{L}_{\rm NRQCD} &=& \psi^\dagger \left\{ i D_0 + \cdots \right\} \psi + \chi^\dagger \left\{ i D_0 + \cdots \right\} \chi
\\ \nonumber &+& {f_1(^1S_0) \over m_1 m_2} \psi^\dagger \chi \chi^\dagger \psi + {f_1(^3S_0) \over m_1 m_2} \psi^\dagger {\boldsymbol \sigma} \chi \chi^\dagger {\boldsymbol \sigma} \psi
\\ \nonumber &+& {f_8(^1S_0) \over m_1 m_2} \psi^\dagger T^a\chi \chi^\dagger T^a \psi
\\ &+& {f_8(^3S_0) \over m_1 m_2} \psi^\dagger T^a {\boldsymbol \sigma} \chi \chi^\dagger T^a {\boldsymbol \sigma} \psi
+ \cdots \, .
\end{eqnarray}
We refer to Ref.~\cite{Brambilla:2004jw} for detailed review of this method. The above Lagrangian contains four four-fermion operators, which can be used to study the annihilation width of a heavy quarkonium into light particles. In this method the Fierz rearrangement is applied to decouple the Dirac and color indices that connect the short-distance part to the long-distance part~\cite{Bodwin:1994jh}.

Compared with this, the quark-based effective field theory for the multiquark system is much more difficult~\cite{Brambilla:2019esw}. Let us attempt to do this for the $Z_c(3900)$. Based on Eq.~(\ref{eq:j}) we can write down an eight-quark operator (the same argument applies for other Lagrangians containing $\eta^{\mathcal Z}_\mu$):
\begin{eqnarray}
\mathcal{L} &=& c_0 \times \eta^{\mathcal Z}_\mu \times \left(\eta^{{\mathcal Z},\mu}\right)^\dagger
\\ \nonumber &=& c_0 \times \left( \epsilon^{abe} \epsilon^{cde}~q_a^T\mathbb{C}\gamma_\mu c_b ~ \bar q_c \gamma_5\mathbb{C}\bar c_d^T - \{ \gamma_\mu \leftrightarrow \gamma_5 \} \right)
\\ \nonumber &\times& \left( \epsilon^{a^\prime b^\prime e^\prime } \epsilon^{c^\prime d^\prime e^\prime }~\bar c_{b^\prime} \gamma^\mu \mathbb{C} \bar q_{a^\prime}^T  ~ c_{d^\prime}^T \mathbb{C} \gamma_5 q_{c^\prime} - \{ \gamma_\mu \leftrightarrow \gamma_5 \} \right) \, ,
\end{eqnarray}
where $c_0$ is a constant. Then we can use the Fierz rearrangement to transform it to be
\begin{eqnarray}
\mathcal{L} &=& c_0 \times \Big( + {1\over3} ~ \bar c_{a} \gamma_5 c_a ~ \bar q_{b} \gamma_\mu q_b
\label{eq:example}
\\
&& \nonumber ~~~~~~~~ - {1\over3} ~ \bar c_{a} \gamma_\mu c_a ~ \bar q_{b} \gamma_5 q_b
\\
&& \nonumber ~~~~~~~~ + {i\over3} ~ \bar c_{a} \gamma^\nu \gamma_5 c_a ~ \bar q_{b} \sigma_{\mu\nu} q_b
\\
&& \nonumber ~~~~~~~~ - {i\over3} ~ \bar c_{a} \sigma_{\mu\nu} c_a ~ \bar q_{b} \gamma^\nu \gamma_5 q_b
\\
&& \nonumber ~~~~~~~~ - {1\over4} ~ {\lambda^n_{ab}}{\lambda^n_{cd}} ~ \bar c_{a} \gamma_5 c_b ~ \bar q_{c} \gamma_\mu q_d
\\
&& \nonumber ~~~~~~~~ + {1\over4} ~ {\lambda^n_{ab}}{\lambda^n_{cd}} ~ \bar c_{a} \gamma_\mu c_b ~ \bar q_{c} \gamma_5 q_d
\\
&& \nonumber ~~~~~~~~ - {i\over4} ~ {\lambda^n_{ab}}{\lambda^n_{cd}} ~ \bar c_{a} \gamma^\nu \gamma_5 c_b ~ \bar q_{c} \sigma_{\mu\nu} q_d
\\
&& \nonumber ~~~~~~~~ + {i\over4} ~ {\lambda^n_{ab}}{\lambda^n_{cd}} ~ \bar c_{a} \sigma_{\mu\nu} c_b ~ \bar q_{c} \gamma^\nu \gamma_5 q_d \Big)
\\ \nonumber &\times& \left( \epsilon^{a^\prime b^\prime e^\prime } \epsilon^{c^\prime d^\prime e^\prime }~\bar c_{b^\prime} \gamma^\mu \mathbb{C} \bar q_{a^\prime}^T  ~ c_{d^\prime}^T \mathbb{C} \gamma_5 q_{c^\prime}
- \{ \gamma_\mu \leftrightarrow \gamma_5 \} \right) \, .
\end{eqnarray}
Detailed discussions on this transformation will be given below.

Considering that the meson operators, $\bar q \gamma_5 q$, $\bar q \gamma_\mu q$, $\bar c \gamma_5 c$, and $\bar c \gamma_\mu c$ couple to the $\pi$, $\rho$, $\eta_c$, and $J/\psi$ mesons respectively (see Table~\ref{tab:coupling} below), the above eight-quark operator can describe the fall-apart decays of the $Z_c(3900)$ into the $\eta_c \rho$ and $J/\psi \pi$ final states simultaneously, together with some other possible decay channels. In order to extract the widths of these decays, one still needs to do further calculations, which we shall not study any more. However, their relative branching ratios can be extracted much more easily, which are also useful and important to understand the nature of the $Z_c(3900)$~\cite{Yu:2017zst}.

The current $\eta^{\mathcal Z}_\mu$ can also be investigated from the viewpoint of QCD sum rules~\cite{Shifman:1978bx,Reinders:1984sr}. We assume it couples to the $Z_c(3900)$ through
\begin{equation}
\langle 0 | \eta^{\mathcal Z}_\mu | Z_c \rangle = f_{Z_c} \epsilon_\mu \, .
\end{equation}
After the Fierz rearrangement, $\eta^{\mathcal Z}_\mu$ transforms to the long expression inside Eq.~(\ref{eq:example}). Through the first and second terms, it couples to the $\eta_c \rho$ and $J/\psi \pi$ channels simultaneously:
\begin{eqnarray}
\langle 0 | \eta^{\mathcal Z}_\mu | \eta_c \rho \rangle &=& {1\over3} \langle 0 | \bar c_{a} \gamma_5 c_a | \eta_c \rangle \langle 0 | \bar q_{b} \gamma_\mu q_b | \rho \rangle + \cdots \, ,
\\ \nonumber \langle 0 | \eta^{\mathcal Z}_\mu | J/\psi \pi \rangle &=& - {1\over3} \langle 0 | \bar c_{a} \gamma_\mu c_a | J/\psi \rangle \langle 0 | \bar q_{b} \gamma_5 q_b | \pi \rangle + \cdots \, .
\end{eqnarray}
Again, these two equations can be easily used to calculate the relative branching ratio ${\mathcal R}_{Z_c}$. Detailed discussions on this will be given below.

In the above equations we have worked within the naive factorization scheme, so our uncertainty is significantly larger than the well-developed QCD factorization method~\cite{Beneke:1999br,Beneke:2000ry,Beneke:2001ev}, which has been widely and successfully applied to study weak and radiative decay properties of conventional (heavy) hadrons, {e.g.}, see Refs.~\cite{Wang:2017ijn,Li:2020rcg}. However, given that we still do not well understand the internal structure of the $Z_c(3900)$ (as well as all the other exotic hadrons), the naive factorization scheme at this moment can be useful. Besides, the tetraquark decay constant $f_{Z_c}$ is removed when calculating relative branching ratios, which significantly reduces our uncertainty.

In the present study we shall study strong decay properties of the $Z_c(3900)$ under the naive factorization scheme. To do this we just need to replace the weak-interaction Lagrangian by some interpolating current, and apply the similar technics here together with the Fierz arrangement. Note that a similar arrangement of the spin and color indices in the nonrelativistic case was used to study strong decay properties of the $Z_c(3900)$ in Refs.~\cite{Voloshin:2013dpa,Maiani:2017kyi,Voloshin:2018pqn}.

This paper is organized as follows. In Sec.~\ref{sec:current} we systematically construct all the tetraquark currents of $J^{PC}=1^{+-}$ with the quark content $c \bar c q \bar q$. There are three configurations, $[cq][\bar c \bar q]$, $[\bar c q][\bar q c]$, and $[\bar c c][\bar q q]$, and their relations are also derived in this section by using the Fierz rearrangement of the Dirac and color indices. In Sec.~\ref{sec:mesoncurrent} we discuss the couplings of meson operators to meson states, and list those which are needed in the present study. In Sec.~\ref{sec:decaydiquark} and Sec.~\ref{sec:decaymolecule} we extract some decay properties of the $Z_c(3900)$, separately for the compact tetraquark interpretation and the hadronic molecule interpretation. The obtained results are discussed and summarized in Sec.~\ref{sec:summary}.

\section{Tetraquark currents of $J^{PC} = 1^{+-}$ and their relations}
\label{sec:current}

\begin{figure*}[hbt]
\begin{center}
\subfigure[~\mbox{$[cq][\bar c \bar q]$ currents $\eta_\mu^i(x,y)$}]{\includegraphics[width=0.23\textwidth]{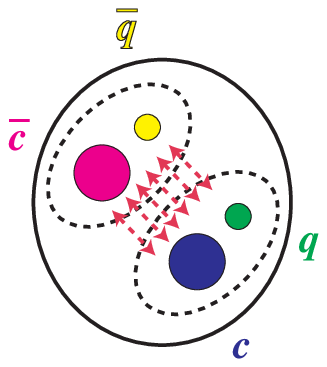}}
~~~~~~
\subfigure[~\mbox{$[\bar c q][\bar q c]$ currents $\xi_\mu^i(x,y)$}]{\includegraphics[width=0.23\textwidth]{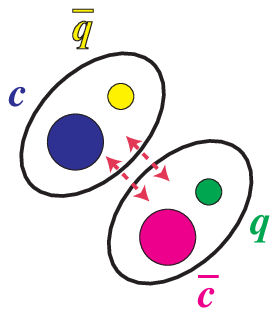}}
~~~~~~
\subfigure[~\mbox{$[\bar c c][\bar q q]$ currents $\theta_\mu^i(x,y)$}]{\includegraphics[width=0.23\textwidth]{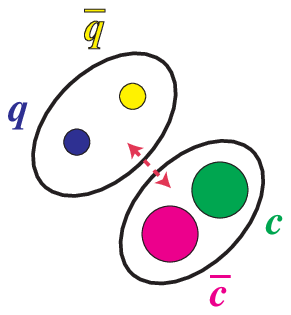}}
\caption{Three types of tetraquark currents. Quarks are shown in red/green/blue color, and antiquarks are shown in cyan/magenta/yellow color.}
\label{fig:current}
\end{center}
\end{figure*}

By using the $c$, $\bar c$, $q$, $\bar q$ quarks ($q=u/d$), one can construct three types of tetraquark currents, as illustrated in Fig.~\ref{fig:current}:
\begin{eqnarray}
\nonumber \eta(x,y)    &=& [q^T_a(x)~\mathbb{C} \Gamma_1~c_b(x)]   \times   [\bar q_c(y)~\Gamma_2 \mathbb{C}~\bar c_d^T(y)] \, ,
\\ \xi(x,y)  &=& [\bar c_a(x)~\Gamma_3~q_b(x)]           \times   [\bar q_c(y)~\Gamma_4~c_d(y)] \, ,
\\ \nonumber \theta(x,y) &=& [\bar c_a(x)~\Gamma_5~c_b(x)]           \times   [\bar q_c(y)~\Gamma_6~q_d(y)] \, ,
\end{eqnarray}
where $\Gamma_i$ are Dirac matrices, $\mathbb{C}$ is the charge-conjugation matrix, the subscripts $a, b, c, d$ are color indices, and the sum over repeated indices is taken.
One usually call $\eta(x,y)$ the diquark-antidiquark current, and $\xi(x,y)$ and $\theta(x,y)$ the mesonic-mesonic currents. We separately construct them as follows.

%
%

\subsection{$[qc][\bar q \bar c]$ currents $\eta_\mu^i(x,y)$}

There are altogether eight independent $[qc][\bar q \bar c]$ currents of $J^{PC} = 1^{+-}$~\cite{Chen:2010ze}:
\begin{eqnarray}
\eta^1_\mu &=& q_a^T\mathbb{C}\gamma_\mu c_b ~ \bar q_{a} \gamma_5\mathbb{C}\bar c_{b}^T - q_a^T\mathbb{C}\gamma_5 c_b ~ \bar q_{a} \gamma_\mu\mathbb{C}\bar c_{b}^T \, ,
\\
\nonumber \eta^2_\mu &=& q_a^T\mathbb{C}\gamma_\mu c_b ~ \bar q_{b} \gamma_5\mathbb{C}\bar c_{a}^T - q_a^T\mathbb{C}\gamma_5 c_b ~ \bar q_{b} \gamma_\mu\mathbb{C}\bar c_{a}^T \, ,
\\
\nonumber \eta^3_\mu &=& q_a^T\mathbb{C}\gamma^\nu c_b ~ \bar q_{a} \sigma_{\mu\nu} \gamma_5\mathbb{C}\bar c_{b}^T - q_a^T\mathbb{C}\sigma_{\mu\nu} \gamma_5 c_b ~ \bar q_{a} \gamma^\nu\mathbb{C}\bar c_{b}^T \, ,
\\
\nonumber \eta^4_\mu &=& q_a^T\mathbb{C}\gamma^\nu c_b ~ \bar q_{b} \sigma_{\mu\nu} \gamma_5\mathbb{C}\bar c_{a}^T - q_a^T\mathbb{C}\sigma_{\mu\nu} \gamma_5 c_b ~ \bar q_{b} \gamma^\nu\mathbb{C}\bar c_{a}^T \, ,
\\
\nonumber \eta^5_\mu &=& q_a^T\mathbb{C}\gamma_\mu \gamma_5 c_b ~ \bar q_{a}\mathbb{C}\bar c_{b}^T - q_a^T\mathbb{C}c_b ~ \bar q_{a} \gamma_\mu \gamma_5\mathbb{C}\bar c_{b}^T \, ,
\\
\nonumber \eta^6_\mu &=& q_a^T\mathbb{C}\gamma_\mu \gamma_5 c_b ~ \bar q_{b}\mathbb{C}\bar c_{a}^T - q_a^T\mathbb{C}c_b ~ \bar q_{b} \gamma_\mu \gamma_5\mathbb{C}\bar c_{a}^T \, ,
\\
\nonumber \eta^7_\mu &=& q_a^T\mathbb{C}\gamma^\nu \gamma_5 c_b ~ \bar q_{a} \sigma_{\mu\nu}\mathbb{C}\bar c_{b}^T - q_a^T\mathbb{C}\sigma_{\mu\nu} c_b ~ \bar q_{a} \gamma^\nu \gamma_5\mathbb{C}\bar c_{b}^T \, ,
\\
\nonumber \eta^8_\mu &=& q_a^T\mathbb{C}\gamma^\nu \gamma_5 c_b ~ \bar q_{b} \sigma_{\mu\nu}\mathbb{C}\bar c_{a}^T - q_a^T\mathbb{C}\sigma_{\mu\nu} c_b ~ \bar q_{b} \gamma^\nu \gamma_5\mathbb{C}\bar c_{a}^T \, .
\end{eqnarray}
Here we have omitted the coordinates $x$ and $y$ for simplicity. Their combinations, $\eta^1_\mu - \eta^2_\mu$, $\eta^3_\mu - \eta^4_\mu$, $\eta^5_\mu - \eta^6_\mu$, and $\eta^7_\mu - \eta^8_\mu$ have the antisymmetric color structure $[q c]_{\mathbf{\bar 3}_c}[\bar q \bar c]_{\mathbf{3}_c}\rightarrow[c \bar c q \bar q]_{\mathbf{1}_c}$, and $\eta^1_\mu + \eta^2_\mu$, $\eta^3_\mu + \eta^4_\mu$, $\eta^5_\mu + \eta^6_\mu$, and $\eta^7_\mu + \eta^8_\mu$ have the symmetric color structure $[q c]_{\mathbf{6}_c}[\bar q \bar c]_{\mathbf{\bar 6}_c}\rightarrow[c \bar c q \bar q]_{\mathbf{1}_c}$.

In the ``type-II'' diquark-antidiquark model proposed in Ref.~\cite{Maiani:2014aja}, the ground-state tetraquarks can be written in the spin basis as $|s_{qc}, s_{\bar q \bar c} \rangle_J$, where $s_{qc}$ and $s_{\bar q \bar c}$ are the charmed diquark and antidiquark spins, respectively. There are two ground-state diquarks: the ``good'' one of $J^P = 0^+$ and the ``bad'' one of $J^P = 1^+$~\cite{Jaffe:2004ph}. By combining them, the $Z_c(3900)$ was interpreted as a diquark-antidiquark state of $J^{PC} = 1^{+-}$ in Ref.~\cite{Maiani:2014aja}:
\begin{equation}
|0_{qc}1_{\bar q \bar c}; 1^{+-} \rangle = {1\over\sqrt2} \left(| 0_{qc}, 1_{\bar q \bar c} \rangle_{J=1} - |1_{qc}, 0_{\bar q \bar c} \rangle_{J=1} \right) \, .
\label{eq:diquarkZ}
\end{equation}
The interpolating current having the identical internal structure is just the current $\eta^{\mathcal Z}_\mu$ given in Eq.~(\ref{eq:j}), which has been studied in Ref.~\cite{Dias:2013xfa,Agaev:2016dev,Wang:2017lot,Chen:2019osl} using QCD sum rules:
\begin{eqnarray}
\eta^{\mathcal Z}_\mu(x,y) &=& \eta^1_\mu([uc][\bar d \bar c]) - \eta^2_\mu([uc][\bar d \bar c])
\label{eq:diquark}
\\ \nonumber &=& u_a^T(x)\mathbb{C}\gamma_\mu c_b(x)  \left( \bar d_{a}(y) \gamma_5\mathbb{C}\bar c_{b}^T(y) - \{ a \leftrightarrow b \} \right)
\\ \nonumber &&  - ~\{ \gamma_\mu \leftrightarrow \gamma_5 \} \, .
\end{eqnarray}
Here we have explicitly chosen the quark content $[uc][\bar d \bar c]$ for the positive-charged one $Z_c(3900)^+$.

\subsection{$[\bar c q][\bar q c]$ currents $\xi_\mu^i(x,y)$}

There are altogether eight independent $[\bar c q][\bar q c]$ currents of $J^{PC} = 1^{+-}$:
\begin{eqnarray}
\xi^1_\mu &=& \bar c_{a} \gamma_\mu q_a ~ \bar q_{b} \gamma_5 c_b + \bar c_{a} \gamma_5 q_a ~ \bar q_{b} \gamma_\mu c_b \, ,
\\
\nonumber \xi^2_\mu &=& \bar c_{a} \gamma^\nu q_a ~ \bar q_{b} \sigma_{\mu\nu} \gamma_5 c_b - \bar c_{a} \sigma_{\mu\nu} \gamma_5 q_a ~ \bar q_{b} \gamma^\nu c_b \, ,
\\
\nonumber \xi^3_\mu &=& \bar c_{a} \gamma_\mu \gamma_5 q_a ~ \bar q_{b} c_b - \bar c_{a} q_a ~ \bar q_{b} \gamma_\mu \gamma_5 c_b \, ,
\\
\nonumber \xi^4_\mu &=& \bar c_{a} \gamma^\nu \gamma_5 q_a ~ \bar q_{b} \sigma_{\mu\nu} c_b + \bar c_{a} \sigma_{\mu\nu} q_a ~ \bar q_{b} \gamma^\nu \gamma_5 c_b \, ,
\\
\nonumber \xi^5_\mu &=& {\lambda^n_{ab}}{\lambda^n_{cd}} ~ \left( \bar c_{a} \gamma_\mu q_b ~ \bar q_{c} \gamma_5 c_d + \bar c_{a} \gamma_5 q_b ~ \bar q_{c} \gamma_\mu c_d \right) \, ,
\\
\nonumber \xi^6_\mu &=& {\lambda^n_{ab}}{\lambda^n_{cd}} ~ \left( \bar c_{a} \gamma^\nu q_b ~ \bar q_{c} \sigma_{\mu\nu} \gamma_5 c_d - \bar c_{a} \sigma_{\mu\nu} \gamma_5 q_b ~ \bar q_{c} \gamma^\nu c_d \right) \, ,
\\
\nonumber \xi^7_\mu &=& {\lambda^n_{ab}}{\lambda^n_{cd}} ~ \left( \bar c_{a} \gamma_\mu \gamma_5 q_b ~ \bar q_{c} c_d - \bar c_{a} q_b ~ \bar q_{c} \gamma_\mu \gamma_5 c_d \right) \, ,
\\
\nonumber \xi^8_\mu &=& {\lambda^n_{ab}}{\lambda^n_{cd}} ~ \left( \bar c_{a} \gamma^\nu \gamma_5 q_b ~ \bar q_{c} \sigma_{\mu\nu} c_d + \bar c_{a} \sigma_{\mu\nu} q_b ~ \bar q_{c} \gamma^\nu \gamma_5 c_d \right) \, .
\end{eqnarray}
Among them, $\xi^{1,2,3,4}_\mu$ have the color structure $[\bar c q]_{\mathbf{1}_c}[\bar q c]_{\mathbf{1}_c}\rightarrow[c \bar c q \bar q]_{\mathbf{1}_c}$, and $\xi^{5,6,7,8}_\mu$ have the color structure $[\bar c q]_{\mathbf{8}_c}[\bar q c]_{\mathbf{8}_c}\rightarrow[c \bar c q \bar q]_{\mathbf{1}_c}$. In the molecular picture the $Z_c(3900)$ can be interpreted as the $D \bar D^*$ hadronic molecular state of $J^{PC} = 1^{+-}$~\cite{Voloshin:1976ap,Voloshin:2013dpa,Wang:2013cya,Guo:2013sya}:
\begin{equation}
| D \bar D^*; 1^{+-} \rangle = {1\over\sqrt2} \left(| D \bar D^* \rangle_{J=1} - | \bar D D^* \rangle_{J=1} \right)  \, ,
\label{eq:moleculeZ}
\end{equation}
and the relevant interpolating current is~\cite{Cui:2013yva,Zhang:2013aoa,Chen:2015ata}:
\begin{eqnarray}
\xi^{\mathcal Z}_\mu(x,y) &=& \xi^1_\mu([\bar c u][\bar d c])
\label{eq:molecule}
\\ \nonumber &=& \bar c_{a}(x) \gamma_\mu u_a(x) ~ \bar d_{b}(y) \gamma_5 c_b(y) + \{ \gamma_\mu \leftrightarrow \gamma_5 \} \, .
\end{eqnarray}
Again, we have chosen the quark content $[\bar c u][\bar d c]$.

\subsection{$[\bar c c][\bar q q]$ currents $\theta_\mu^i(x,y)$}

There are altogether eight independent $[\bar c c][\bar q q]$ currents of $J^{PC} = 1^{+-}$:
\begin{eqnarray}
\theta^1_\mu(x,y) &=& \bar c_{a}(x) \gamma_5 c_a(x) ~ \bar q_{b}(y) \gamma_\mu q_b(y) \, ,
\\
\nonumber \theta^2_\mu(x,y) &=& \bar c_{a}(x) \gamma_\mu c_a(x) ~ \bar q_{b}(y) \gamma_5 q_b(y) \, ,
\\
\nonumber \theta^3_\mu(x,y) &=& \bar c_{a}(x) \gamma^\nu \gamma_5 c_a(x) ~ \bar q_{b}(y) \sigma_{\mu\nu} q_b(y) \, ,
\\
\nonumber \theta^4_\mu(x,y) &=& \bar c_{a}(x) \sigma_{\mu\nu} c_a(x) ~ \bar q_{b}(y) \gamma^\nu \gamma_5 q_b(y) \, ,
\\
\nonumber \theta^5_\mu(x,y) &=& {\lambda^n_{ab}}{\lambda^n_{cd}} ~ \bar c_{a}(x) \gamma_5 c_b(x) ~ \bar q_{c}(y) \gamma_\mu q_d(y) \, ,
\\
\nonumber \theta^6_\mu(x,y) &=& {\lambda^n_{ab}}{\lambda^n_{cd}} ~ \bar c_{a}(x) \gamma_\mu c_b(x) ~ \bar q_{c}(y) \gamma_5 q_d(y) \, ,
\\
\nonumber \theta^7_\mu(x,y) &=& {\lambda^n_{ab}}{\lambda^n_{cd}} ~ \bar c_{a}(x) \gamma^\nu \gamma_5 c_b(x) ~ \bar q_{c}(y) \sigma_{\mu\nu} q_d(y) \, ,
\\
\nonumber \theta^8_\mu(x,y) &=& {\lambda^n_{ab}}{\lambda^n_{cd}} ~ \bar c_{a}(x) \sigma_{\mu\nu} c_b(x) ~ \bar q_{c}(y) \gamma^\nu \gamma_5 q_d(y) \, .
\end{eqnarray}
Among them, $\theta^{1,2,3,4}_\mu$ have the color structure $[\bar c c]_{\mathbf{1}_c}[\bar q q]_{\mathbf{1}_c}\rightarrow[c \bar c q \bar q]_{\mathbf{1}_c}$, and $\theta^{5,6,7,8}_\mu$ have the color structure $[\bar c c]_{\mathbf{8}_c}[\bar q q]_{\mathbf{8}_c}\rightarrow[c \bar c q \bar q]_{\mathbf{1}_c}$. We will discuss their corresponding hadron states in Sec.~\ref{sec:mesoncurrent}.

\subsection{Fierz rearrangement}

We have applied the Fierz rearrangement of the Dirac and color indices to systematically study light baryon and tetraquark operators/currents in Refs.~\cite{Chen:2006hy,Chen:2006zh,Chen:2007xr,Chen:2008ej,Chen:2008qw,Chen:2009tpa,Jiao:2009ra,Chen:2012ut,Chen:2013gnu,Chen:2013jra,Cui:2019roq,Chen:2008qv,Chen:2009sf,Chen:2012ex}. It can also be used to relate the above three types of tetraquark currents. To do this, we need to use a) the Fierz transformation~\cite{fierz} in the Lorentz space to rearrange the Dirac indices, and b) the color rearrangement in the color space to rearrange the color indices. All the necessary equations can be found in Sec.~3.3.2 of Ref.~\cite{Chen:2016qju}.

In Eq.~(\ref{eq:example}) the Fierz rearrangement is applied to local operators/currents. However, the Fierz rearrangement is actually a matrix identity, which is valid if the same quark field in the initial and final operators is at the same location. As an example, we can apply the Fierz rearrangement to transform the non-local current with the quark fields $\eta(x,x^\prime;y,y^\prime) = [q(x) c(x^\prime)] [\bar q(y) \bar c(y^\prime)]$ into a combination of several non-local currents with the quark fields at same locations $\xi(y^\prime,x;y,x^\prime) = [\bar c(y^\prime) q(x)] [\bar q(y) c(x^\prime)]$.

\begin{widetext}
Altogether, we obtain the following relation between the currents $\eta^i_\mu(x,x^\prime;y,y^\prime)$ and $\theta^i_\mu(y^\prime,x^\prime;y,x)$:
\begin{equation}
\left(\begin{array}{c}
\eta^1_\mu
\\
\eta^2_\mu
\\
\eta^3_\mu
\\
\eta^4_\mu
\\
\eta^5_\mu
\\
\eta^6_\mu
\\
\eta^7_\mu
\\
\eta^8_\mu
\end{array}\right)
=
\left(\begin{array}{cccccccc}
{1/2} & -{1/2} & {i/2} & -{i/2} & 0         & 0          & 0         & 0
\\
{1/6} & -{1/6} & {i/6} & -{i/6} & {1/4} & -{1/4} & {i/4} & -{i/4}
\\
{3i/2} & {3i/2} & -{1/2} & -{1/2} & {0} & {0} & {0} & {0}
\\
{i/2} & {i/2} & -{1/6} & -{1/6} & {3i/4} & {3i/4} & -{1/4} & -{1/4}
\\
{1/2} & {1/2} & -{i/2} & -{i/2} & {0} & {0} & {0} & {0}
\\
{1/6} & {1/6} & -{i/6} & -{i/6} & {1/4} & {1/4} & -{i/4} & -{i/4}
\\
{3i/2} & -{3i/2} & {1/2} & -{1/2} & {0} & {0} & {0} & {0}
\\
{i/2} & -{i/2} & {1/6} & -{1/6} & {3i/4} & -{3i/4} & {1/4} & -{1/4}
\end{array}\right)
\times
\left(\begin{array}{c}
\theta^1_\mu
\\
\theta^2_\mu
\\
\theta^3_\mu
\\
\theta^4_\mu
\\
\theta^5_\mu
\\
\theta^6_\mu
\\
\theta^7_\mu
\\
\theta^8_\mu
\end{array}\right) \, ,
\label{eq:fierz1}
\end{equation}
the following relation between $\eta^i_\mu(x,x^\prime;y,y^\prime)$ and $\xi^i_\mu(y^\prime,x;y,x^\prime)$:
\begin{equation}
\left(\begin{array}{c}
\eta^1_\mu
\\
\eta^2_\mu
\\
\eta^3_\mu
\\
\eta^4_\mu
\\
\eta^5_\mu
\\
\eta^6_\mu
\\
\eta^7_\mu
\\
\eta^8_\mu
\end{array}\right)
=
\left(\begin{array}{cccccccc}
0 & {i/6} & -{1/6} & 0 & 0 & {i/4} & -{1/4} & 0
\\
0 & {i/2} & -{1/2} & 0 & 0 & 0 & 0 & 0
\\
-{i/2} & 0 & 0 & {1/6} & -{3i/4} & {0} & {0} & {1/4}
\\
-{3i/2} & 0 & 0 & {1/2} & 0 & 0 & 0 & 0
\\
{1/6} & 0 & 0 & -{i/6} & {1/4} & {0} & {0} & -{i/4}
\\
{1/2} & 0 & 0 & -{i/2} & 0 & 0 & 0 & 0
\\
0 & -{1/6} & {i/2} & 0 & {0} & -{1/4} & {3i/4} & {0}
\\
0 & -{1/2} & {3i/2} & 0 & 0 & 0 & 0 & 0
\end{array}\right)
\times
\left(\begin{array}{c}
\xi^1_\mu
\\
\xi^2_\mu
\\
\xi^3_\mu
\\
\xi^4_\mu
\\
\xi^5_\mu
\\
\xi^6_\mu
\\
\xi^7_\mu
\\
\xi^8_\mu
\end{array}\right) \, ,
\label{eq:fierz2}
\end{equation}
the following relation among $\eta^i_\mu(x,x^\prime;y,y^\prime)$, $\xi^{1,2,3,4}_\mu(y^\prime,x;y,x^\prime)$, and $\theta^{1,2,3,4}_\mu(y^\prime,x^\prime;y,x)$:
\begin{equation}
\left(\begin{array}{c}
\eta^1_\mu
\\
\eta^2_\mu
\\
\eta^3_\mu
\\
\eta^4_\mu
\\
\eta^5_\mu
\\
\eta^6_\mu
\\
\eta^7_\mu
\\
\eta^8_\mu
\end{array}\right)
=
\left(\begin{array}{cccccccc}
0 & 0 & 0 & 0 & {1/2} & -{1/2} & {i/2} & -{i/2}
\\
0 & {i/2} & -{1/2} & 0 & 0 & 0 & 0 & 0
\\
0 & 0 & 0 & 0 & {3i/2} & {3i/2} & -{1/2} & -{1/2}
\\
-{3i/2} & 0 & 0 & {1/2} & 0 & 0 & 0 & 0
\\
0 & 0 & 0 & 0 & {1/2} & {1/2} & -{i/2} & -{i/2}
\\
{1/2} & 0 & 0 & -{i/2} & 0 & 0 & 0 & 0
\\
0 & 0 & 0 & 0 & {3i/2} & -{3i/2} & {1/2} & -{1/2}
\\
0 & -{1/2} & {3i/2} & 0 & 0 & 0 & 0 & 0
\end{array}\right)
\times
\left(\begin{array}{c}
\xi^1_\mu
\\
\xi^2_\mu
\\
\xi^3_\mu
\\
\xi^4_\mu
\\
\theta^1_\mu
\\
\theta^2_\mu
\\
\theta^3_\mu
\\
\theta^4_\mu
\end{array}\right) \, ,
\label{eq:fierz3}
\end{equation}
and the following relation between $\xi^i_\mu(y^\prime,x;y,x^\prime)$ and $\theta^i_\mu(y^\prime,x^\prime;y,x)$:
\begin{equation}
\left(\begin{array}{c}
\xi^1_\mu
\\
\xi^2_\mu
\\
\xi^3_\mu
\\
\xi^4_\mu
\\
\xi^5_\mu
\\
\xi^6_\mu
\\
\xi^7_\mu
\\
\xi^8_\mu
\end{array}\right)
=
\left(\begin{array}{cccccccc}
-{1/6} & -{1/6} & -{i/6} & -{i/6} & -{1/4} & -{1/4} & -{i/4} & -{i/4}
\\
-{i/2} & {i/2} & {1/6} & -{1/6} & -{3i/4} & {3i/4} & {1/4} & -{1/4}
\\
{1/6} & -{1/6} & -{i/6} & {i/6} & {1/4} & -{1/4} & -{i/4} & {i/4}
\\
{i/2} & {i/2} & {1/6} & {1/6} & {3i/4} & {3i/4} & {1/4} & {1/4}
\\
-{8/9} & -{8/9} & -{8i/9} & -{8i/9} & {1/6} & {1/6} & {i/6} & {i/6}
\\
-{8i/3} & {8i/3} & {8/9} & -{8/9} & {i/2} & -{i/2} & -{1/6} & {1/6}
\\
{8/9} & -{8/9} & -{8i/9} & {8i/9} & -{1/6} & {1/6} & {i/6} & -{i/6}
\\
{8i/3} & {8i/3} & {8/9} & {8/9} & -{i/2} & -{i/2} & -{1/6} & -{1/6}
\end{array}\right)
\times
\left(\begin{array}{c}
\theta^1_\mu
\\
\theta^2_\mu
\\
\theta^3_\mu
\\
\theta^4_\mu
\\
\theta^5_\mu
\\
\theta^6_\mu
\\
\theta^7_\mu
\\
\theta^8_\mu
\end{array}\right) \, .
\label{eq:fierz4}
\end{equation}
\end{widetext}

\section{Meson operators}
\label{sec:mesoncurrent}

\begin{table*}[hbt]
\begin{center}
\renewcommand{\arraystretch}{1.5}
\caption{Couplings of meson operators to meson states. Color indices are omitted for simplicity.}
\begin{tabular}{ c | c | c | c | c | c}
\hline\hline
~~~Operators~~~ & ~~~$J^{PC}$~~~ & ~~~Mesons~~~ & ~~~$J^{PC}$~~~ & ~~~Couplings~~~ & ~~~Decay Constants~~~
\\ \hline\hline
$J^{S} = \bar d u$ & $0^{++}$ & -- & $0^{++}$ & -- & --
\\ \hline
$J^{P} = \bar d i\gamma_5 u$ & $0^{-+}$ &  $\pi^+$  & $0^{-+}$ &  $\langle 0 | J^{P} | \pi^+ \rangle = \lambda_\pi$  &  $\lambda_\pi = {f_\pi m_\pi^2 \over m_u + m_d}$
\\ \hline
$J^{V}_\mu = \bar d \gamma_\mu u$ & $1^{--}$ &  $\rho^+$  & $1^{--}$ &  $\langle 0 | J^{V}_\mu | \rho^+ \rangle = m_\rho f_{\rho^+} \epsilon_\mu$  &  $f_{\rho^+} = 208$~MeV~\mbox{\cite{Jansen:2009hr}}
\\ \hline
\multirow{2}{*}{$J^{A}_\mu = \bar d \gamma_\mu \gamma_5 u$} & \multirow{2}{*}{$1^{++}$} & $\pi^+$  & $0^{-+}$ & $\langle 0 | J^{A}_\mu | \pi^+\rangle = i p_\mu f_{\pi^+}$  &  $f_{\pi^+} = 130.2$~MeV~\mbox{\cite{pdg}}
\\ \cline{3-6}
                                                         & &  $a_1(1260)$  & $1^{++}$ & $\langle 0 | J^{A}_\mu | a_1 \rangle = m_{a_1} f_{a_1} \epsilon_\mu $  &  $f_{a_1} = 254$~MeV~\mbox{\cite{Wingate:1995hy}}
\\ \hline
\multirow{2}{*}{$J^{T}_{\mu\nu} = \bar d \sigma_{\mu\nu} u$} & \multirow{2}{*}{$1^{\pm-}$} &  $\rho^+$ & $1^{--}$   &  $\langle 0 | J^{T}_{\mu\nu} | \rho^+ \rangle = i f^T_{\rho} (p_\mu\epsilon_\nu - p_\nu\epsilon_\mu) $  &  $f_{\rho}^T = 159$~MeV~\mbox{\cite{Jansen:2009hr}}
\\ \cline{3-6}
                                                             & &  $b_1(1235)$  & $1^{+-}$ &  $\langle 0 | J^{T}_{\mu\nu} | b_1 \rangle = i f^T_{b_1} \epsilon_{\mu\nu\alpha\beta} \epsilon^\alpha p^\beta $  &  $f_{b_1}^T = 180$~MeV~\mbox{\cite{Ball:1996tb}}
\\ \hline\hline
$I^{S} = \bar c c$ & $0^{++}$ & $\chi_{c0}(1P)$  & $0^{++}$ &  $\langle 0 | I^S | \chi_{c0} \rangle = m_{\chi_{c0}} f_{\chi_{c0}}$  &  $f_{\chi_{c0}} = 343$~MeV~\mbox{\cite{Veliev:2010gb}}
\\ \hline
$I^{P} = \bar c i\gamma_5 c$ & $0^{-+}$ &  $\eta_c$  & $0^{-+}$ &  $\langle 0 | I^{P} | \eta_c \rangle = \lambda_{\eta_c}$  &  $\lambda_{\eta_c} = {f_{\eta_c} m_{\eta_c}^2 \over 2 m_c}$
\\ \hline
$I^{V}_\mu = \bar c \gamma_\mu c$ & $1^{--}$ &  $J/\psi$  & $1^{--}$ &  $\langle 0 | I^{V}_\mu | J/\psi \rangle = m_{J/\psi} f_{J/\psi} \epsilon_\mu$  &  $f_{J/\psi} = 418$~MeV~\mbox{\cite{Becirevic:2013bsa}}
\\ \hline
\multirow{2}{*}{$I^{A}_\mu = \bar c \gamma_\mu \gamma_5 c$} & \multirow{2}{*}{$1^{++}$} &  $\eta_c$   & $0^{-+}$ &  $\langle 0 | I^{A}_\mu | \eta_c \rangle = i p_\mu f_{\eta_c}$  &  $f_{\eta_c} = 387$~MeV~\mbox{\cite{Becirevic:2013bsa}}
\\ \cline{3-6}
                                                         &&  $\chi_{c1}(1P)$   & $1^{++}$ &  $\langle 0 | I^{A}_\mu | \chi_{c1} \rangle = m_{\chi_{c1}} f_{\chi_{c1}} \epsilon_\mu $  &  $f_{\chi_{c1}} = 335$~MeV~\mbox{\cite{Novikov:1977dq}}
\\ \hline
\multirow{2}{*}{$I^{T}_{\mu\nu} = \bar c \sigma_{\mu\nu} c$} & \multirow{2}{*}{$1^{\pm-}$} &  $J/\psi$   & $1^{--}$   &  $\langle 0 | I^{T}_{\mu\nu} | J/\psi \rangle = i f^T_{J/\psi} (p_\mu\epsilon_\nu - p_\nu\epsilon_\mu) $  &  $f_{J/\psi}^T = 410$~MeV~\mbox{\cite{Becirevic:2013bsa}}
\\ \cline{3-6}
                                                             &&  $h_c(1P)$   & $1^{+-}$ &  $\langle 0 | I^{T}_{\mu\nu} | h_c \rangle = i f^T_{h_c} \epsilon_{\mu\nu\alpha\beta} \epsilon^\alpha p^\beta $  &  $f_{h_c}^T = 235$~MeV~\mbox{\cite{Becirevic:2013bsa}}
\\ \hline\hline
$O^{S} = \bar d c$ & $0^{+}$ &  $D_0^{*+}$  & $0^{+}$ &  $\langle 0 | O^{S} | D_0^{*+} \rangle = m_{D_0^{*}} f_{D_0^{*}}$  &  $f_{D_0^{*}} = 410$~MeV~\mbox{\cite{Narison:2015nxh}}
\\ \hline
$O^{P} = \bar d i\gamma_5 c$ & $0^{-}$ &  $D^+$  & $0^{-}$ &  $\langle 0 | O^{P} | D^+ \rangle = \lambda_D$  &  $\lambda_D = {f_D m_D^2 \over {m_c + m_d}}$
\\ \hline
$O^{V}_\mu = \bar c \gamma_\mu u$ & $1^{-}$ &  $\bar D^{*0}$  & $1^{-}$ &  $\langle 0 | O^{V}_\mu | \bar D^{*0} \rangle = m_{D^*} f_{D^*} \epsilon_\mu$  &  $f_{D^*} = 253$~MeV~\mbox{\cite{Chang:2018aut}}
\\ \hline
\multirow{2}{*}{$O^{A}_\mu = \bar c \gamma_\mu \gamma_5 u$} & \multirow{2}{*}{$1^{+}$} &  $\bar D^{0}$  & $0^{-}$ &  $\langle 0 | O^{A}_\mu | \bar D^{0} \rangle = i p_\mu f_{D}$  &  $f_{D} = 211.9$~MeV~\mbox{\cite{pdg}}
\\ \cline{3-6}
                                                         &&  $D_1$   & $1^{+}$ &  $\langle 0 | O^{A}_\mu | D_1 \rangle = m_{D_1} f_{D_1} \epsilon_\mu $  &  $f_{D_1} = 356$~MeV~\mbox{\cite{Narison:2015nxh}}
\\ \hline
\multirow{2}{*}{$O^{T}_{\mu\nu} = \bar d \sigma_{\mu\nu} c$} & \multirow{2}{*}{$1^{\pm}$} &  $\bar D^{*+}$   & $1^{-}$   &  $\langle 0 | O^{T}_{\mu\nu} | D^{*+} \rangle = i f_{D^*}^T (p_\mu\epsilon_\nu - p_\nu\epsilon_\mu) $  &  $f_{D^*}^T \approx 220$~MeV
\\ \cline{3-6}
                                                         &&  --   & $1^{+}$ &  --  &  --
\\ \hline\hline
\end{tabular}
\label{tab:coupling}
\end{center}
\end{table*}

There are altogether six types of meson operators: $\bar q_a q_a$, $\bar q_a \gamma_5 q_a$, $\bar q_a \gamma_\mu q_a$, $\bar q_a \gamma_\mu \gamma_5 q_a$, $\bar q_a \sigma_{\mu\nu} q_a$, and $\bar q_a \sigma_{\mu\nu} \gamma_5 q_a$. The last two can be related to each other through
\begin{equation}
\sigma_{\mu\nu} \gamma_5 = {i\over2} \epsilon_{\mu\nu\rho\sigma} \sigma^{\rho\sigma} \, .
\end{equation}
The couplings of these operators to meson states are already well understood, {\it i.e.}, some of them have been measured in particle experiments, and some of them have been studied and calculated by various theoretical methods, such as Lattice QCD and QCD sum rules, etc.

In the present study we need the following couplings, as summarized in Table~\ref{tab:coupling}:
\begin{enumerate}

\item The scalar operators $J^{S} = \bar q_a q_a$ and $I^{S} =\bar c_a c_a$ of $J^{PC} = 0^{++}$ couple to scalar mesons. In Ref.~\cite{Veliev:2010gb} the authors used the method of QCD sum rules and extracted the coupling of $I^{S}$ to $\chi_{c0}(1P)$ to be
\begin{equation}
\langle 0 | \bar c_a c_a | \chi_{c0}(p) \rangle = m_{\chi_{c0}} f_{\chi_{c0}} \, ,
\end{equation}
where
\begin{equation}
f_{\chi_{c0}} = 343~{\rm MeV} \, .
\end{equation}
See also discussions in Refs.~\cite{Novikov:1977dq,Eichten:1995ch,Colangelo:2002mj}. The light scalar mesons have a complicated nature~\cite{Pelaez:2015qba}, so we shall not investigate their relevant decay channels in the present study.

\item The pseudoscalar operators $J^{P} = \bar q_a i\gamma_5 q_a$ and $I^{P} = \bar c_a i\gamma_5 c_a$ of $J^{PC} = 0^{-+}$ couple to the pseudoscalar mesons $\pi$ and $\eta_c$, respectively. We can evaluate them through~\cite{Shuryak:1993kg}:
\begin{eqnarray}
\langle 0 | \bar d_a i\gamma_5 u_a | \pi^+(p) \rangle &=& \lambda_{\pi} = {f_{\pi^+} m_{\pi^+}^2 \over m_u + m_d} \, ,
\\ \nonumber \langle 0 | \bar c_a i\gamma_5 c_a | \eta_c(p) \rangle &=& \lambda_{\eta_c} = {f_{\eta_c} m_{\eta_c}^2 \over 2 m_c} \, .
\end{eqnarray}

\item The vector operators $J^{V}_\mu = \bar q_a \gamma_\mu q_a$ and $I^{V}_\mu = \bar c_a \gamma_\mu c_a$ of $J^{PC} = 1^{--}$ couple to the vector mesons $\rho$ and $J/\psi$, respectively. In Refs.~\cite{Jansen:2009hr,Becirevic:2013bsa} the authors used the method of Lattice QCD to obtain
\begin{eqnarray}
\langle 0 | \bar d_a \gamma_\mu u_a | \rho^+(p, \epsilon) \rangle &=& m_{\rho} f_{\rho^+} \epsilon_\mu \, ,
\\ \nonumber \langle 0 | \bar c_a \gamma_\mu c_a | J/\psi(p, \epsilon) \rangle &=& m_{J/\psi} f_{J/\psi} \epsilon_\mu \, ,
\end{eqnarray}
where
\begin{eqnarray}
f_{\rho^+} &=& 208~{\rm MeV} \, ,
\\ \nonumber f_{J/\psi} &=& 418~{\rm MeV} \, .
\end{eqnarray}
See also discussions in Refs.~\cite{Zhu:1998bm,Jiang:2015paa,Colangelo:2000dp}.

\item The axialvector operators $J^{A}_\mu = \bar q_a \gamma_\mu \gamma_5 q_a$ and $I^{A}_\mu = \bar c_a \gamma_\mu \gamma_5 c_a$ of $J^{PC} = 1^{++}$ couple to both pseudoscalar mesons ($\pi$ and $\eta_c$ of $J^{PC} = 0^{-+}$) and axialvector mesons ($a_1(1260)$ and $\chi_{c1}(1P)$ of $J^{PC} = 1^{++}$). The coupling of $J^{A}_\mu$ to $\pi$ has been well measured in particle experiments~\cite{pdg}:
\begin{equation}
\langle 0 | \bar d_a \gamma_\mu \gamma_5 u_a | \pi^+(p) \rangle = i p_\mu f_{\pi^+} \, ,
\end{equation}
while its coupling to $a_1(1260)$ was evaluated by using Lattice QCD~\cite{Wingate:1995hy}:
\begin{equation}
\langle 0 | \bar d_a \gamma_\mu \gamma_5 u_a | a_1(p, \epsilon) \rangle = m_{a_1} f_{a_1} \epsilon_\mu \, ,
\end{equation}
where
\begin{eqnarray}
f_{\pi^+} &=& 130.2~{\rm MeV} \, ,
\\ \nonumber f_{a_1} &=& 254~{\rm MeV} \, .
\end{eqnarray}
The coupling of $I^{A}_\mu$ to $\eta_c$ and $\chi_{c1}(1P)$ was evaluated by using Lattice QCD~\cite{Becirevic:2013bsa} and QCD sum rules~\cite{Novikov:1977dq}:
\begin{eqnarray}
\langle 0 | \bar c_a \gamma_\mu \gamma_5 c_a | \eta_c(p) \rangle &=& i p_\mu f_{\eta_c} \, ,
\\ \nonumber \langle 0 | \bar c_a \gamma_\mu \gamma_5 c_a | \chi_{c1}(p, \epsilon) \rangle &=& m_{\chi_{c1}} f_{\chi_{c1}} \epsilon_\mu \, ,
\end{eqnarray}
where
\begin{eqnarray}
f_{\eta_c} &=& 387~{\rm MeV} \, ,
\\ \nonumber f_{\chi_{c1}} &=& 335~{\rm MeV} \, .
\end{eqnarray}
See also discussions in Refs.~\cite{Simonov:2015cxa,Jiang:2015paa,Yang:2007zt,DiSalvo:1994dg,Braguta:2006wr,Veliev:2010vd,Veliev:2011kq,Novikov:1977dq,Wang:2008qw}.

\item The tensor operators $J^{T}_{\mu\nu} = \bar q_a \sigma_{\mu\nu} q_a$ and $I^{T}_{\mu\nu} = \bar c_a \sigma_{\mu\nu} c_a$ of $J^{PC} = 1^{\pm-}$ couple to both vector mesons ($\rho$ and $J/\psi$ of $J^{PC} = 1^{--}$) and axialvector mesons ($b_1(1235)$ and $h_{c}(1P)$ of $J^{PC} = 1^{+-}$). The coupling of $J^{T}_{\mu\nu}$ to $\rho$ and $b_1(1235)$ was calculated through Lattice QCD~\cite{Jansen:2009hr} and QCD sum rules~\cite{Ball:1996tb}:
\begin{eqnarray}
\langle 0 | \bar d_a \sigma_{\mu\nu} u_a | \rho^+(p, \epsilon) \rangle &=& i f^T_{\rho} (p_\mu\epsilon_\nu - p_\nu\epsilon_\mu) \, ,
\\ \nonumber \langle 0 | \bar d_a \sigma_{\mu\nu} u_a | b_1(p, \epsilon) \rangle &=& i f^T_{b_1} \epsilon_{\mu\nu\alpha\beta} \epsilon^\alpha p^\beta \, ,
\end{eqnarray}
where
\begin{eqnarray}
f_{\rho}^T &=& 159~{\rm MeV} \, ,
\\ \nonumber f_{b_1}^T &=& 180~{\rm MeV} \, .
\end{eqnarray}
The coupling of $I^{T}_{\mu\nu}$ to $J/\psi$ and $h_{c}(1P)$ was calculated through Lattice QCD~\cite{Becirevic:2013bsa}:
\begin{eqnarray}
\langle 0 | \bar c_a \sigma_{\mu\nu} c_a | J/\psi(p, \epsilon) \rangle &=& i f^T_{J/\psi} (p_\mu\epsilon_\nu - p_\nu\epsilon_\mu) \, ,
\\ \nonumber \langle 0 | \bar c_a \sigma_{\mu\nu} c_a | h_c(p, \epsilon) \rangle &=& i f^T_{h_c} \epsilon_{\mu\nu\alpha\beta} \epsilon^\alpha p^\beta \, ,
\end{eqnarray}
where
\begin{eqnarray}
f_{J/\psi}^T &=& 410~{\rm MeV} \, ,
\\ \nonumber f_{h_c}^T &=& 235~{\rm MeV} \, .
\end{eqnarray}
See also discussions in Refs.~\cite{Chizhov:2003qy,Jansen:2009yh,Bakulev:2000er,Yang:2005gk,Ball:1998kk,Becirevic:2004qv,Braun:2003jg,Braguta:2007fh,Wang:2012gj}.

\item The $Z_c(3900)$ is above the $D \bar D^*$ threshold, so we need the couplings of $O^{P} = \bar q_a i\gamma_5 c_a$ and $O^{A}_\mu = \bar c_a \gamma_\mu \gamma_5 q_a$ to the $D$ meson~\cite{pdg}:
\begin{eqnarray}
\langle 0 | \bar d_a i\gamma_5 c_a | D^+(p) \rangle &=& \lambda_D \, ,
\\ \nonumber \langle 0 | \bar c_a \gamma_\mu \gamma_5 u_a | \bar D^0(p) \rangle &=& i p_\mu f_{D} \, ,
\end{eqnarray}
and the couplings of $O^{V}_\mu = \bar c_a \gamma_\mu q_a$ and $O^{T}_{\mu\nu} = \bar q_a \sigma_{\mu\nu} c_a$ to the $D^*$ meson~\cite{Chang:2018aut}:
\begin{eqnarray}
\langle 0 | \bar c_a \gamma_\mu u_a | \bar D^{*0}(p, \epsilon) \rangle &=& m_{D^*} f_{D^*} \epsilon_\mu \, ,
\\ \nonumber \langle 0 | \bar d_a \sigma_{\mu\nu} c_a | D^{*+}(p, \epsilon) \rangle &=& i f^T_{D^*} (p_\mu\epsilon_\nu - p_\nu\epsilon_\mu) \, ,
\end{eqnarray}
where
\begin{eqnarray}
\nonumber \lambda_D &=& {f_{D} m_{D^+}^2 \over {m_c + m_d}} \, ,
\\ f_{D} &=& 211.9~{\rm MeV} \, ,
\\ \nonumber f_{D^*} &=& 253~{\rm MeV} \, .
\end{eqnarray}
We do not find any theoretical study on the transverse decay constant $f^T_{D^*}$, so we simply fit among the decay constants, $f_{\pi^+}$-$f_{\rho^+}$-$f_{\rho}^T$, $f_{\eta_c}$-$f_{J/\psi}$-$f_{J/\psi}^T$, and $f_{D}$-$f_{D^*}$-$f_{D^*}^T$, to obtain
\begin{equation}
f_{D^*}^T \approx 220~{\rm MeV} \, .
\end{equation}
See also discussions in Refs.~\cite{Lubicz:2017asp,Narison:2014ska}.

\item The $Z_c(3900) \rightarrow D \bar D^*_0 \rightarrow D \bar D \pi$ decay is kinematically allowed, so we need the coupling of $O^{S} = \bar q_a c_a$ to the $D_0^*$ meson~\cite{Narison:2015nxh}:
\begin{equation}
\langle 0 | \bar d_a c_a | D_0^{*+}(p) \rangle = m_{D_0^{*}} f_{D_0^{*}} \, ,
\end{equation}
where
\begin{equation}
f_{D_0^{*}} = 410~{\rm MeV} \, .
\end{equation}
See also discussions in Refs.~\cite{Herdoiza:2006qv,Wang:2015mxa}.

\end{enumerate}

\section{Decay properties of the $Z_c(3900)$ as a compact tetraquark state}
\label{sec:decaydiquark}

In this section and the next we shall use Eqs.~(\ref{eq:fierz1}-\ref{eq:fierz4}) derived in Sec.~\ref{sec:current} to extract some decay properties of the $Z_c(3900)$. The two possible interpretations of the $Z_c(3900)$ are: a) the compact tetraquark state of $J^{PC} = 1^{+-}$ composed of a $J^P= 0^+$ diquark/antidiquark and a $J^P= 1^+$ antidiquark/diquark~\cite{Maiani:2004vq,Maiani:2014aja}, {\it i.e.}, $|0_{qc}1_{\bar q \bar c} ; 1^{+-} \rangle$ defined in Eq.~(\ref{eq:diquarkZ}); and b) the $D \bar D^*$ hadronic molecular state of $J^{PC} = 1^{+-}$~\cite{Voloshin:1976ap,Voloshin:2013dpa,Wang:2013cya,Guo:2013sya}, {\it i.e.}, $| D \bar D^*; 1^{+-} \rangle$ defined in Eq.~(\ref{eq:moleculeZ}).
Moreover, we shall study their mixing with the $|1_{qc}1_{\bar q \bar c} ; 1^{+-} \rangle$ and $| D^* \bar D^*; 1^{+-} \rangle$ states, whose definitions will be given below.

In this section we shall investigate the former compact tetraquark interpretation, whose relevant current $\eta^{\mathcal Z}_\mu(x,y)$ has been given in Eq.~(\ref{eq:diquark}). This current can be transformed to $\theta_\mu^i(x,y)$ and $\xi_\mu^i(x,y)$ according to Eqs.~(\ref{eq:fierz1}-\ref{eq:fierz3}), through which we shall extract some decay properties of the $Z_c(3900)$ as a compact tetraquark state in the following subsections.

\subsection{$\eta^{\mathcal Z}_\mu\big([uc][\bar d \bar c]\big) \rightarrow \theta_\mu^i \big([\bar c c] + [\bar d u]\big)$}
\label{sec:diquark1}

%
\begin{figure*}[hbt]
\begin{center}
\includegraphics[width=0.8\textwidth]{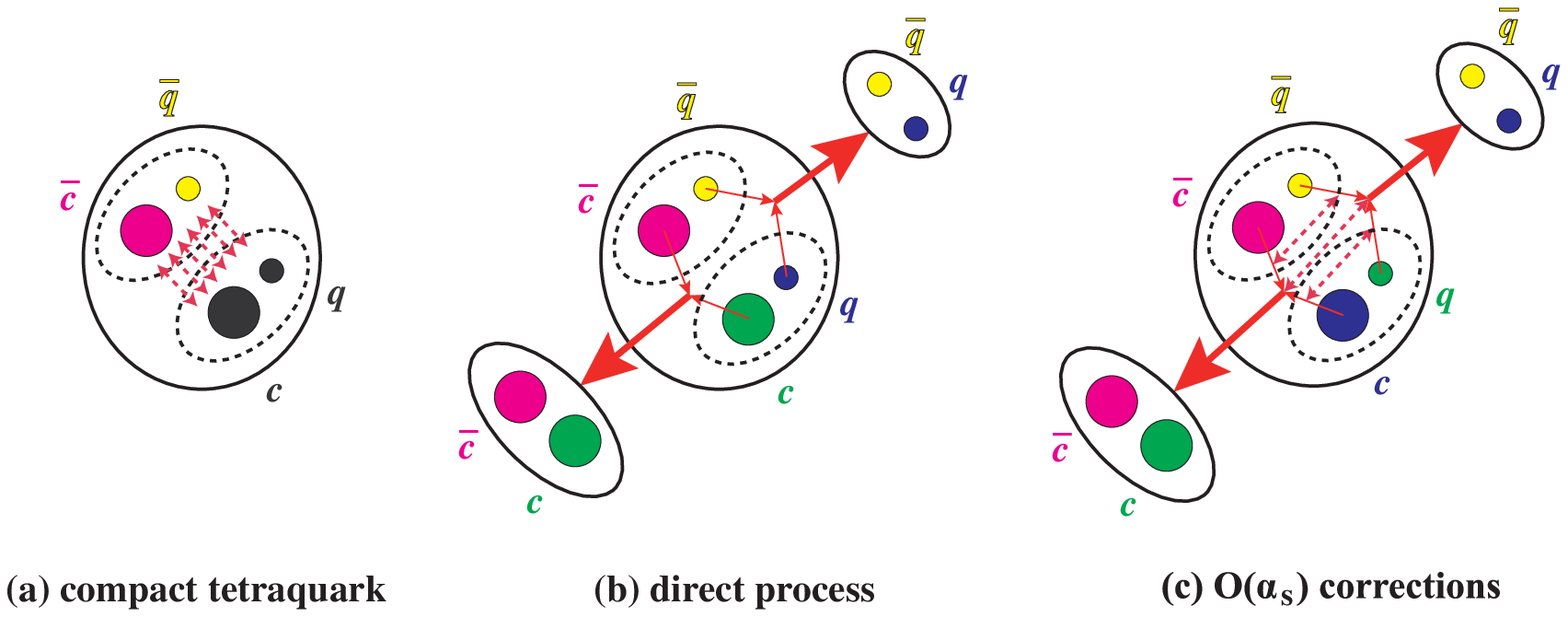}
\caption{The decay of a compact tetraquark (diquark-antidiquark) state into one charmonium meson and one light meson. This decay can happen through either (b) a direct fall-apart process, or (c) a process with gluon(s) exchanged, that is the $\mathcal{O}(\alpha_s)$ corrections.}
\label{fig:a}
\end{center}
\end{figure*}
%

As depicted in Fig.~\ref{fig:a}, when the $c$ and $\bar c$ quarks meet each other and the $u$ and $\bar d$ quarks meet each other at the same time, a compact tetraquark state can decay into one charmonium meson and one light meson:
\begin{eqnarray}
&& [u(x) c(x)]~[\bar d(y) \bar c(y)]
\label{eq:change1}
\\ \nonumber &\Longrightarrow& [u(x \to y^\prime)~c(x \to x^\prime)]~[\bar d(y \to y^\prime)~\bar c(y \to x^\prime)]
\\ \nonumber &\Longrightarrow& [\bar c(x^\prime) c(x^\prime)] + [\bar d(y^\prime) u(y^\prime)] \, .
\end{eqnarray}
The first process is a dynamical process, during which we assume that all the flavor, color, spin and orbital structures remain unchanged, so the relevant current also remains the same. The second process for $|0_{qc}1_{\bar q \bar c} ; 1^{+-} \rangle$ can be described by the transformation~(\ref{eq:fierz1}):
\begin{eqnarray}
&& \eta^{\mathcal Z}_\mu(x,y)
\label{tran:diquark1}
\\ \nonumber &\Longrightarrow& + {1\over3}~\theta_\mu^1(x^\prime,y^\prime) - {1\over3}~\theta_\mu^2(x^\prime,y^\prime)
\\ \nonumber && + {i\over3}~\theta_\mu^3(x^\prime,y^\prime) - {i\over3}~\theta_\mu^4(x^\prime,y^\prime) + \cdots
\\ \nonumber &=& - {i\over3}~I^{P}(x^\prime) ~ J^{V}_\mu(y^\prime) + {i\over3}~I^{V}_\mu(x^\prime) ~ J^{P}(y^\prime)
\\ \nonumber &&  + {i\over3}~I^{A,\nu}(x^\prime) ~ J^{T}_{\mu\nu}(y^\prime) - {i\over3}~I^{T}_{\mu\nu}(x^\prime) ~ J^{A,\nu}(y^\prime) + \cdots \, ,
\end{eqnarray}
where we have only kept the direct fall-apart process described by $\theta_\mu^{1,2,3,4}$, but neglected the $\mathcal{O}(\alpha_s)$ corrections described by $\theta_\mu^{5,6,7,8}$.

Together with Table~\ref{tab:coupling}, we extract the following decay channels from the above transformation:
\begin{enumerate}

\item The decay of $|0_{qc}1_{\bar q \bar c} ; 1^{+-} \rangle$ into $\eta_c\rho$ is contributed by both $I^{P} \times J^{V}_\mu$ and $I^{A,\nu} \times J^{T}_{\mu\nu}$:
\begin{eqnarray}
&& \langle Z_c^+(p,\epsilon) | \eta_c(p_1)~\rho^+(p_2,\epsilon_2) \rangle
\label{decay:etacrho}
\\ \nonumber &\approx& - {i c_1\over3}~\lambda_{\eta_c} m_\rho f_{\rho^+}~\epsilon \cdot \epsilon_2
\\ \nonumber && ~~~~~ - {i c_1\over3}~f_{\eta_c} f^T_{\rho}~(\epsilon \cdot p_2 ~ \epsilon_2 \cdot p_1 - p_1 \cdot p_2~\epsilon \cdot \epsilon_2)
\\ \nonumber &\equiv& g^S_{\eta_c \rho}~\epsilon \cdot \epsilon_2 + g^D_{\eta_c \rho}~(\epsilon \cdot p_2 ~ \epsilon_2 \cdot p_1 - p_1 \cdot p_2~\epsilon \cdot \epsilon_2) \, ,
\end{eqnarray}
where $c_1$ is an overall factor, related to the coupling of $\eta^{\mathcal Z}_\mu(x,y)$ to the $Z_c(3900)^+$ as well as the dynamical process $(x,y) \Longrightarrow (x^\prime, y^\prime)$ shown in Fig.~\ref{fig:a}. The two coupling constants $g^S_{\eta_c \rho}$ and $g^D_{\eta_c \rho}$ are defined for the $S$- and $D$-wave $Z_c(3900) \to \eta_c \rho$ decays:
\begin{eqnarray}
\mathcal{L}^S_{\eta_c \rho} &=& g^S_{\eta_c \rho}~Z_{c}^{+,\mu}~\eta_c~\rho^{-}_{\mu} + \cdots \, ,
\label{lag:etcrhoS}
\\ \mathcal{L}^D_{\eta_c \rho} &=& g^D_{\eta_c \rho} \times \left( g^{\mu\sigma}g^{\nu\rho} - g^{\mu\nu}g^{\rho\sigma} \right)
\label{lag:etcrhoD}
\\ \nonumber && ~~~~~ \times Z_{c,\mu}^{+}~\partial_\rho \eta_c~\partial_\sigma \rho^{-}_{\nu} + \cdots \, .
\end{eqnarray}

\item The decay of $|0_{qc}1_{\bar q \bar c} ; 1^{+-} \rangle$ into $J/\psi \pi$ is contributed by both $I^{V}_\mu \times J^{P}$ and $I^{T}_{\mu\nu} \times J^{A,\nu}$:
\begin{eqnarray}
&& \langle Z_c^+(p,\epsilon) | J/\psi(p_1,\epsilon_1)~\pi^+(p_2) \rangle
\label{decay:psipi}
\\ \nonumber &\approx& {i c_1 \over3}~\lambda_{\pi} m_{J/\psi} f_{J/\psi}~\epsilon \cdot \epsilon_1
\\ \nonumber && ~~~~~ + {i c_1 \over3}~f_{\pi^+} f^T_{J/\psi}~(\epsilon \cdot p_1 ~ \epsilon_1 \cdot p_2 - p_1 \cdot p_2~\epsilon \cdot \epsilon_1)
\\ \nonumber &\equiv& g^S_{\psi \pi}~\epsilon \cdot \epsilon_1 + g^D_{\psi \pi}~(\epsilon \cdot p_1 ~ \epsilon_1 \cdot p_2 - p_1 \cdot p_2~\epsilon \cdot \epsilon_1) \, .
\end{eqnarray}
The two coupling constants $g^S_{\psi \pi}$ and $g^D_{\psi \pi}$ are defined for the $S$- and $D$-wave $Z_c(3900) \rightarrow J/\psi \pi$ decays respectively:
\begin{eqnarray}
\mathcal{L}^S_{\psi \pi} &=& g^S_{\psi \pi}~Z_{c}^{+,\mu}~\psi_\mu~\pi^- + \cdots \, ,
\label{lag:psipiS}
\\ \mathcal{L}^D_{\psi \pi} &=& g^D_{\psi \pi} \times \left( g^{\mu\rho}g^{\nu\sigma} - g^{\mu\nu}g^{\rho\sigma} \right)
\label{lag:psipiD}
\\ \nonumber && ~~~~~ \times Z_{c,\mu}^{+}~\partial_\rho \psi_\nu~\partial_\sigma \pi^- + \cdots \, .
\end{eqnarray}

\item The decay of $|0_{qc}1_{\bar q \bar c} ; 1^{+-} \rangle$ into $\eta_c b_1$ is contributed by $I^{A,\nu} \times J^{T}_{\mu\nu}$:
\begin{eqnarray}
&& \langle Z_c^+(p,\epsilon) | \eta_c(p_1)~b_1^+(p_2,\epsilon_2) \rangle
\\ \nonumber &\approx& - {i c_1 \over3}~f_{\eta_c} f^T_{b_1}~\epsilon_{\mu\nu\alpha\beta} \epsilon^\mu p_1^\nu \epsilon_2^\alpha p_2^\beta
\\ \nonumber &\equiv& g_{\eta_c b_1}~\epsilon_{\mu\nu\alpha\beta} \epsilon^\mu p_1^\nu \epsilon_2^\alpha p_2^\beta \, .
\end{eqnarray}
This process is kinematically forbidden, but the $|0_{qc}1_{\bar q \bar c} ; 1^{+-} \rangle \rightarrow \eta_c b_1 \rightarrow \eta_c \omega \pi \rightarrow \eta_c + 4 \pi$ decay is kinematically allowed.

\item The decay of $|0_{qc}1_{\bar q \bar c} ; 1^{+-} \rangle$ into $\chi_{c1} \rho$ is contributed by $I^{A,\nu} \times J^{T}_{\mu\nu}$:
\begin{eqnarray}
&&\langle Z_c^+(p,\epsilon) | \chi_{c1}(p_1,\epsilon_1)~\rho^+(p_2,\epsilon_2) \rangle
\\ \nonumber &\approx& - {c_1\over3}~m_{\chi_{c1}} f_{\chi_{c1}} f^T_{\rho}~(\epsilon_1 \cdot \epsilon_2~\epsilon \cdot p_2 - \epsilon_1 \cdot p_2~\epsilon \cdot \epsilon_2)
\\ \nonumber &\equiv& g_{\chi_{c1} \rho}~(\epsilon_1 \cdot \epsilon_2~\epsilon \cdot p_2 - \epsilon_1 \cdot p_2~\epsilon \cdot \epsilon_2) \, .
\end{eqnarray}
This process is kinematically forbidden, but the $|0_{qc}1_{\bar q \bar c} ; 1^{+-} \rangle \rightarrow \chi_{c1} \rho \rightarrow \chi_{c1} \pi \pi$ decay is kinematically allowed.

\item The decay of $|0_{qc}1_{\bar q \bar c} ; 1^{+-} \rangle$ into $\chi_{c1} b_1$ is contributed by $I^{A,\nu} \times J^{T}_{\mu\nu}$:
\begin{eqnarray}
&& \langle Z_c^+(p,\epsilon) | \chi_{c1}(p_1,\epsilon_1)~b_1^+(p_2,\epsilon_2) \rangle
\\ \nonumber &\approx& - {c_1\over3}~m_{\chi_{c1}} f_{\chi_{c1}} f^T_{b_1}~\epsilon_{\mu\nu\alpha\beta} \epsilon^\mu \epsilon_1^\nu \epsilon_2^\alpha p_2^\beta
\\ \nonumber &\equiv& g_{\chi_{c1} b_1}~\epsilon_{\mu\nu\alpha\beta} \epsilon^\mu \epsilon_1^\nu \epsilon_2^\alpha p_2^\beta \, .
\end{eqnarray}
This process is kinematically forbidden.

\item The decay of $|0_{qc}1_{\bar q \bar c} ; 1^{+-} \rangle$ into $h_c \pi$ is contributed by $I^{T}_{\mu\nu} \times J^{A,\nu}$:
\begin{eqnarray}
&& \langle Z_c^+(p,\epsilon) | h_c(p_1,\epsilon_1)~\pi^+(p_2) \rangle
\\ \nonumber &\approx& {i c_1\over3}~f_{\pi^+} f^T_{h_c}~\epsilon_{\mu\nu\alpha\beta} \epsilon^\mu p_2^\nu \epsilon_1^\alpha p_1^\beta
\\ \nonumber &\equiv& g_{h_c \pi}~\epsilon_{\mu\nu\alpha\beta} \epsilon^\mu p_2^\nu \epsilon_1^\alpha p_1^\beta \, .
\end{eqnarray}
This process is kinematically allowed.

\item The decay of $|0_{qc}1_{\bar q \bar c} ; 1^{+-} \rangle$ into $J/\psi a_1$ is contributed by $I^{T}_{\mu\nu} \times J^{A,\nu}$:
\begin{eqnarray}
&& \langle Z_c^+(p,\epsilon) | J/\psi(p_1,\epsilon_1)~a_1^+(p_2,\epsilon_2) \rangle
\\ \nonumber &\approx& {c_1\over3}~f^T_{J/\psi} m_{a_1} f_{a_1}~(\epsilon_1 \cdot \epsilon_2~\epsilon \cdot p_1 - \epsilon_2 \cdot p_1~ \epsilon \cdot \epsilon_1)
\\ \nonumber &\equiv& g_{\psi a_1}~(\epsilon_1 \cdot \epsilon_2~\epsilon \cdot p_1 - \epsilon_2 \cdot p_1~ \epsilon \cdot \epsilon_1) \, .
\end{eqnarray}
This process is kinematically forbidden, but the $|0_{qc}1_{\bar q \bar c} ; 1^{+-} \rangle \rightarrow J/\psi a_1 \rightarrow J/\psi \rho \pi \rightarrow J/\psi + 3 \pi$ decay is kinematically allowed.

\item The decay of $|0_{qc}1_{\bar q \bar c} ; 1^{+-} \rangle$ into $h_c a_1$ is contributed by $I^{T}_{\mu\nu} \times J^{A,\nu}$:
\begin{eqnarray}
&& \langle Z_c^+(p,\epsilon) | h_c(p_1,\epsilon_1)~a_1^+(p_2,\epsilon_2) \rangle
\\ \nonumber &\approx& {c_1\over3}~f^T_{h_c} m_{a_1} f_{a_1}~\epsilon_{\mu\nu\alpha\beta} \epsilon^\mu \epsilon_2^\nu \epsilon_1^\alpha p_1^\beta
\\ \nonumber &\equiv& g_{h_c a_1}~\epsilon_{\mu\nu\alpha\beta} \epsilon^\mu \epsilon_2^\nu \epsilon_1^\alpha p_1^\beta \, .
\end{eqnarray}
This process is kinematically forbidden.

\end{enumerate}
Summarizing the above results, we obtain numerically
\begin{eqnarray}
\nonumber g^S_{\eta_c \rho} &=& - i c_1~7.29 \times 10^{10}~{\rm MeV}^4 \, ,
\\ \nonumber g^D_{\eta_c \rho} &=& - i c_1~2.05 \times 10^{4}~{\rm MeV}^2 \, ,
\\ \nonumber g^S_{\psi \pi} &=& i c_1~11.87 \times 10^{10}~{\rm MeV}^4 \, ,
\\ \nonumber g^D_{\psi \pi} &=& i c_1~1.78 \times 10^{4}~{\rm MeV}^2 \, ,
\\ g_{\eta_{c} b_1} &=& - i c_1~2.32 \times 10^{4}~{\rm MeV}^2 \, ,
\\ \nonumber g_{\chi_{c1} \rho} &=& - c_1~6.23 \times 10^{7}~{\rm MeV}^3 \, ,
\\ \nonumber g_{\chi_{c1} b_1} &=& - c_1~7.06 \times 10^{7}~{\rm MeV}^3 \, ,
\\ \nonumber g_{h_c \pi} &=& i c_1~1.02 \times 10^{4}~{\rm MeV}^2 \, ,
\\ \nonumber g_{\psi a_1} &=& c_1~4.27 \times 10^{7}~{\rm MeV}^3 \, ,
\\ \nonumber g_{h_c a_1} &=& c_1~2.45 \times 10^{7}~{\rm MeV}^3 \, .
\end{eqnarray}
From these coupling constants, we further obtain the following relative branching ratios, which are kinematically allowed:
\begin{eqnarray}
\nonumber {\mathcal{B}(|0_{qc}1_{\bar q \bar c} ; 1^{+-} \rangle \rightarrow \eta_c\rho) \over \mathcal{B}(|0_{qc}1_{\bar q \bar c} ; 1^{+-} \rangle \rightarrow J/\psi\pi)} &=& 0.059 \, ,
\label{eq:diquarkbr1}
\\ {\mathcal{B}(|0_{qc}1_{\bar q \bar c} ; 1^{+-} \rangle \rightarrow h_c\pi) \over \mathcal{B}(|0_{qc}1_{\bar q \bar c} ; 1^{+-} \rangle \rightarrow J/\psi\pi)} &=& 0.0088 \, ,
\\ \nonumber {\mathcal{B}(|0_{qc}1_{\bar q \bar c} ; 1^{+-} \rangle \rightarrow \chi_{c1}\rho \rightarrow \chi_{c1}\pi \pi) \over \mathcal{B}(|0_{qc}1_{\bar q \bar c} ; 1^{+-} \rangle \rightarrow J/\psi\pi)} &=& 1.4 \times 10^{-6} \, .
\end{eqnarray}
Besides them, the following decay chains are also possible but with quite small partial decay widths:
\begin{eqnarray}
&& |0_{qc}1_{\bar q \bar c} ; 1^{+-} \rangle \rightarrow \eta_c b_1 \rightarrow \eta_c \omega \pi \rightarrow \eta_c + 4 \pi \, ,
\\ \nonumber && |0_{qc}1_{\bar q \bar c} ; 1^{+-} \rangle \rightarrow J/\psi a_1 \rightarrow J/\psi \rho \pi \rightarrow J/\psi + 3 \pi \, .
\end{eqnarray}

\subsection{$\eta^{\mathcal Z}_\mu\big([uc][\bar d \bar c]\big) \rightarrow \xi_\mu^i \big([\bar c u] + [\bar d c]\big)$}
\label{sec:diquark2}

%
\begin{figure*}[hbt]
\begin{center}
\includegraphics[width=0.8\textwidth]{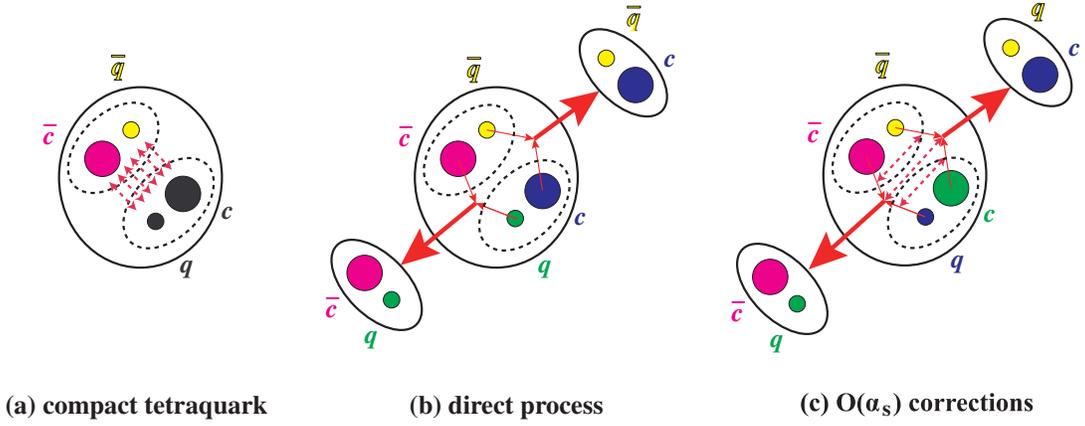}
\caption{The decay of a compact tetraquark (diquark-antidiquark) state into two charmed mesons. This decay can happen through either (b) a direct fall-apart process, or (c) a process with gluon(s) exchanged, that is the $\mathcal{O}(\alpha_s)$ corrections.}
\label{fig:b}
\end{center}
\end{figure*}
%

As depicted in Fig.~\ref{fig:b}, when the $c$ and $\bar d$ quarks meet each other and the $u$ and $\bar c$ quarks meet each other at the same time, a compact tetraquark state can decay into two charmed mesons. This process for $|0_{qc}1_{\bar q \bar c} ; 1^{+-} \rangle$ can be described by the transformation~(\ref{eq:fierz2}):
\begin{equation}
\eta^{\mathcal Z}_\mu(x,y) \Longrightarrow - {i\over3}~\xi_\mu^2(x^\prime,y^\prime) + {1\over3}~\xi_\mu^3(x^\prime,y^\prime) + \cdots \, .
\label{tran:diquark2}
\end{equation}
Again, we have only kept the direct fall-apart process described by $\xi_\mu^{2,3}$, but neglected the $\mathcal{O}(\alpha_s)$ corrections described by $\xi_\mu^{6,7}$.

The term $\xi_\mu^{2}$ couples to the $D^*\bar D^*$ and $D^* \bar D_1$ final states, and the term $\xi_\mu^{3}$ couples to the $D \bar D_0^*$ and $D_1 \bar D_0^*$ final states. Among them, only the $|0_{qc}1_{\bar q \bar c} ; 1^{+-} \rangle \rightarrow D \bar D^*_0 \rightarrow D \bar D \pi$ decay is kinematically allowed, contributed by $\xi_\mu^{3} = O^{A}_\mu \times O^{S}$ to be:
\begin{eqnarray}
\nonumber \langle Z_c^+(p,\epsilon) | \bar D^0(p_1)D_0^{*+}(p_2) \rangle &\approx& {i c_2\over3}~f_D m_{D_0^*} f_{D_0^*}~\epsilon \cdot p_1
\\ &\equiv& g_{D \bar D_0^*}~\epsilon \cdot p_1 \, ,
\\ \nonumber \langle Z_c^+(p,\epsilon) | D^+(p_1)\bar D_0^{*0}(p_2) \rangle &\approx& - {i c_2\over3}~f_D m_{D_0^*} f_{D_0^*}~\epsilon \cdot p_1
\\ &\equiv& - g_{D \bar D_0^*}~\epsilon \cdot p_1 \, ,
\end{eqnarray}
where $c_2$ is an overall factor.

Numerically, we obtain
\begin{equation}
g_{D \bar D_0^*} = ic_2~6.80 \times 10^{7}~{\rm MeV}^3 \, .
\end{equation}
Comparing the $|0_{qc}1_{\bar q \bar c} ; 1^{+-} \rangle \rightarrow D \bar D^*_0 \rightarrow D \bar D \pi$ decay studied in the present subsection with the $|0_{qc}1_{\bar q \bar c} ; 1^{+-} \rangle \rightarrow J/\psi \pi$ and $|0_{qc}1_{\bar q \bar c} ; 1^{+-} \rangle \rightarrow \eta_c\rho$ decays studied in the previous subsection, we obtain
\begin{eqnarray}
\nonumber && {\mathcal{B}(|0_{qc}1_{\bar q \bar c} ; 1^{+-} \rangle \rightarrow D \bar D_0^{*} + \bar D D_0^{*} \rightarrow D \bar D \pi) \over \mathcal{B}(|0_{qc}1_{\bar q \bar c} ; 1^{+-} \rangle \rightarrow J/\psi\pi+ \eta_c\rho)}
\\ &=& 9.3 \times 10^{-8} \times {c_2^2 \over c_1^2} \, .
\label{eq:diquarkDD1}
\end{eqnarray}

The current $\eta^{\mathcal Z}_\mu(x,y)$ does not correlate with the two terms $\xi_\mu^{1}  = -i O^{V}_\mu \times O^{P}$ and $\xi_\mu^{4} = O^{A,\nu} \times O^{T}_{\mu\nu}$, both of which can couple to the $D \bar D^*$ final state. This suggests that $|0_{qc}1_{\bar q \bar c} ; 1^{+-} \rangle$ does not decay to the $D \bar D^*$ final state with a large branching ratio,
\begin{equation}
g_{D \bar D^{*}} \approx 0 \, ,
\end{equation}
so that
\begin{equation}
{\mathcal{B}(|0_{qc}1_{\bar q \bar c} ; 1^{+-} \rangle \rightarrow D \bar D^{*} + \bar D D^{*}) \over \mathcal{B}(|0_{qc}1_{\bar q \bar c} ; 1^{+-} \rangle \rightarrow J/\psi\pi + \eta_c\rho)} \approx 0 \, .
\label{eq:diquarkDD2}
\end{equation}
Eqs.~(\ref{eq:diquarkDD1}) and (\ref{eq:diquarkDD2}) together suggest that $|0_{qc}1_{\bar q \bar c} ; 1^{+-} \rangle$ mainly decays into one charmonium meson and one light meson, other than two charmed mesons.

\subsection{$\eta^{\mathcal Z}_\mu\big([uc][\bar d \bar c]\big) \rightarrow \theta_\mu^{1,2,3,4}\big([\bar c c] + [\bar d u]\big) + \xi_\mu^{1,2,3,4}\big([\bar c u] + [\bar d c]\big)$}
\label{sec:diquark3}

If the above two processes investigated in Sec.~\ref{sec:diquark1} and Sec.~\ref{sec:diquark2} happen at the same time, we can use the transformation~(\ref{eq:fierz3}), {\it i.e.}, $|0_{qc}1_{\bar q \bar c} ; 1^{+-} \rangle$ can decay into one charmonium meson and one light meson as well as two charmed mesons at the same time, which process is described by the color-singlet-color-singlet currents $\theta_\mu^{1,2,3,4}$ and $\xi_\mu^{1,2,3,4}$ together:
\begin{eqnarray}
&& \eta^{\mathcal Z}_\mu(x,y)
\label{tran:diquark3}
\\ \nonumber &\Longrightarrow& + {1\over2}~\theta_\mu^1(x^\prime,y^\prime) - {1\over2}~\theta_\mu^2(x^\prime,y^\prime) + {i\over2}~\theta_\mu^3(x^\prime,y^\prime)
\\ \nonumber && - {i\over2}~\theta_\mu^4(x^\prime,y^\prime) - {i\over2}~\xi_\mu^2(x^{\prime\prime},y^{\prime\prime}) + {1\over2}~\xi_\mu^3(x^{\prime\prime},y^{\prime\prime}) \, .
\end{eqnarray}
Here we have kept all the terms, and there is no $\cdots$ in this equation.

Comparing the above equation with Eqs.~(\ref{tran:diquark1}) and (\ref{tran:diquark2}), we obtain the same relative branching ratios as Sec.~\ref{sec:diquark1} and Sec.~\ref{sec:diquark2}, just with the overall factors $c_1$ and $c_2$ replaced by others.

\subsection{Mixing with $|1_{qc}1_{\bar q \bar c} ; 1^{+-} \rangle$}
\label{sec:diquarkmixing}

The relative branching ratio ${\mathcal R}_{Z_c}$ calculated in Sec.~\ref{sec:diquark1} is just $0.059$, significantly smaller than the BESIII measurement ${\mathcal R}_{Z_c} = 2.2 \pm 0.9$ at $\sqrt{s} = 4.226$~GeV~\cite{Ablikim:2019ipd}. In this subsection we slightly modify the internal structure of the $Z_c(3900)$ to reevaluate this ratio.

Actually, in the Type-II diquark-antidiquark model~\cite{Maiani:2014aja}, the $Z_c(3900)$ was interpreted as
\begin{equation*}
|0_{qc}1_{\bar q \bar c} ; 1^{+-} \rangle = {1\over\sqrt2} \left(| 0_{qc}, 1_{\bar q \bar c} \rangle_{J=1} - |1_{qc}, 0_{\bar q \bar c} \rangle_{J=1} \right) \, ,
\end{equation*}
and the ratio ${\mathcal R}_{Z_c}$ was predicted to be $0.27^{+0.40}_{-0.17}$~\cite{Esposito:2014hsa}; while in the Type-I diquark-antidiquark model~\cite{Maiani:2004vq}, the $Z_c(3900)$ was interpreted as the mixing state
\begin{equation}
|x_{qc}1_{\bar q \bar c} ; 1^{+-} \rangle = \cos \theta_1 ~ |0_{qc}1_{\bar q \bar c} ; 1^{+-} \rangle + \sin \theta_1 ~ |1_{qc}1_{\bar q \bar c} ; 1^{+-} \rangle \, ,
\label{eq:diquarkmixing}
\end{equation}
where
\begin{equation}
|1_{qc}1_{\bar q \bar c} ; 1^{+-} \rangle = | 1_{qc}, 1_{\bar q \bar c} \rangle_{J=1} \, ,
\label{eq:diquarkZp}
\end{equation}
and a small $|1_{qc}1_{\bar q \bar c} ; 1^{+-} \rangle$ component is able to increase this ratio to be $\left( 2.3^{+3.3}_{-1.4} \right) \times 10^2$~\cite{Esposito:2014hsa}, that is almost one thousand times larger.

\begin{figure*}[htb]
\begin{center}
\subfigure[]{\scalebox{0.48}{\includegraphics{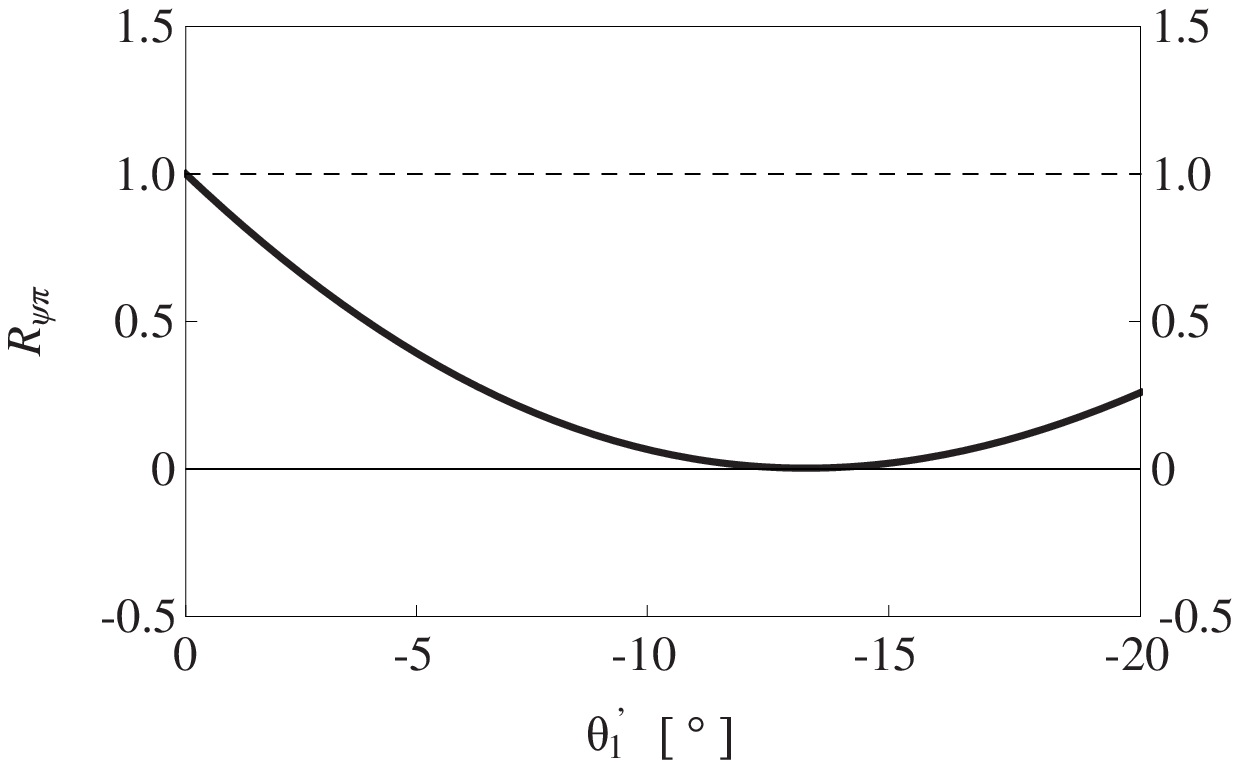}}}~
\subfigure[]{\scalebox{0.445}{\includegraphics{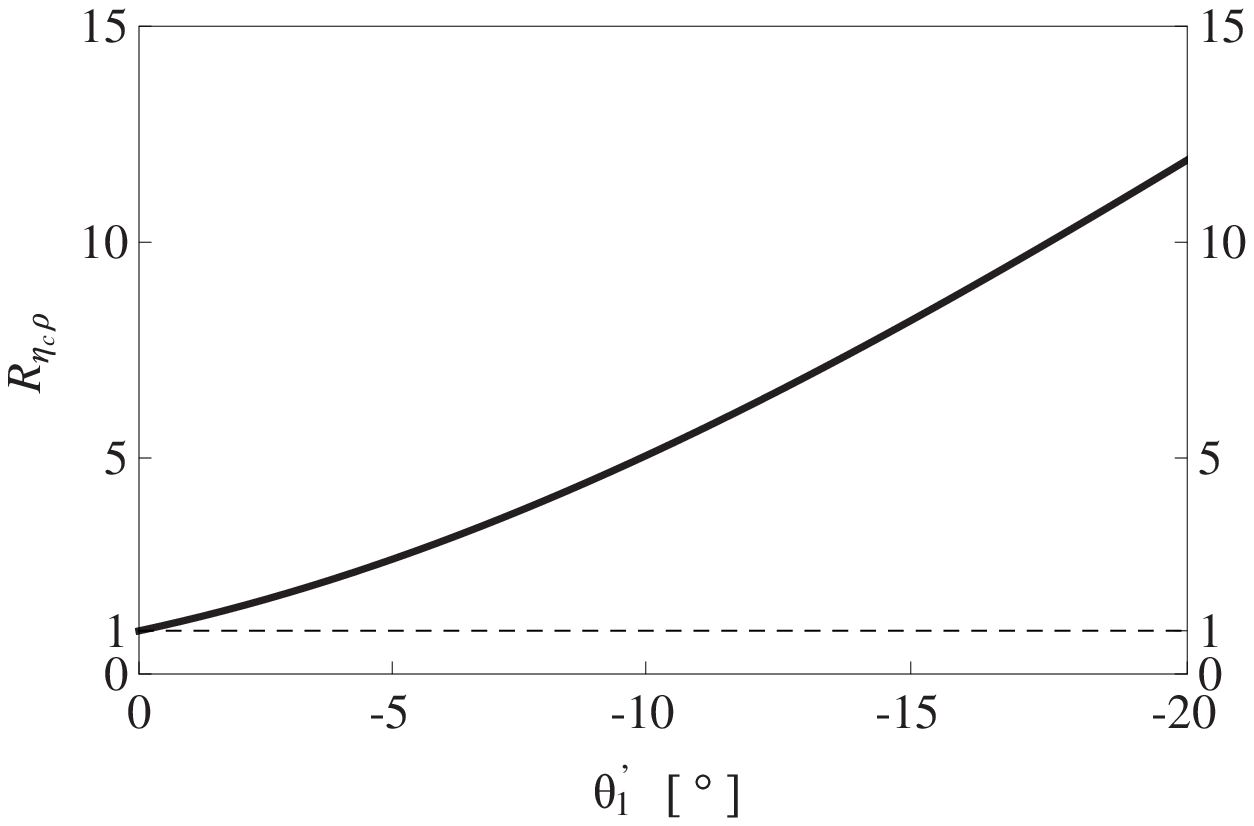}}}~
\subfigure[]{\scalebox{0.452}{\includegraphics{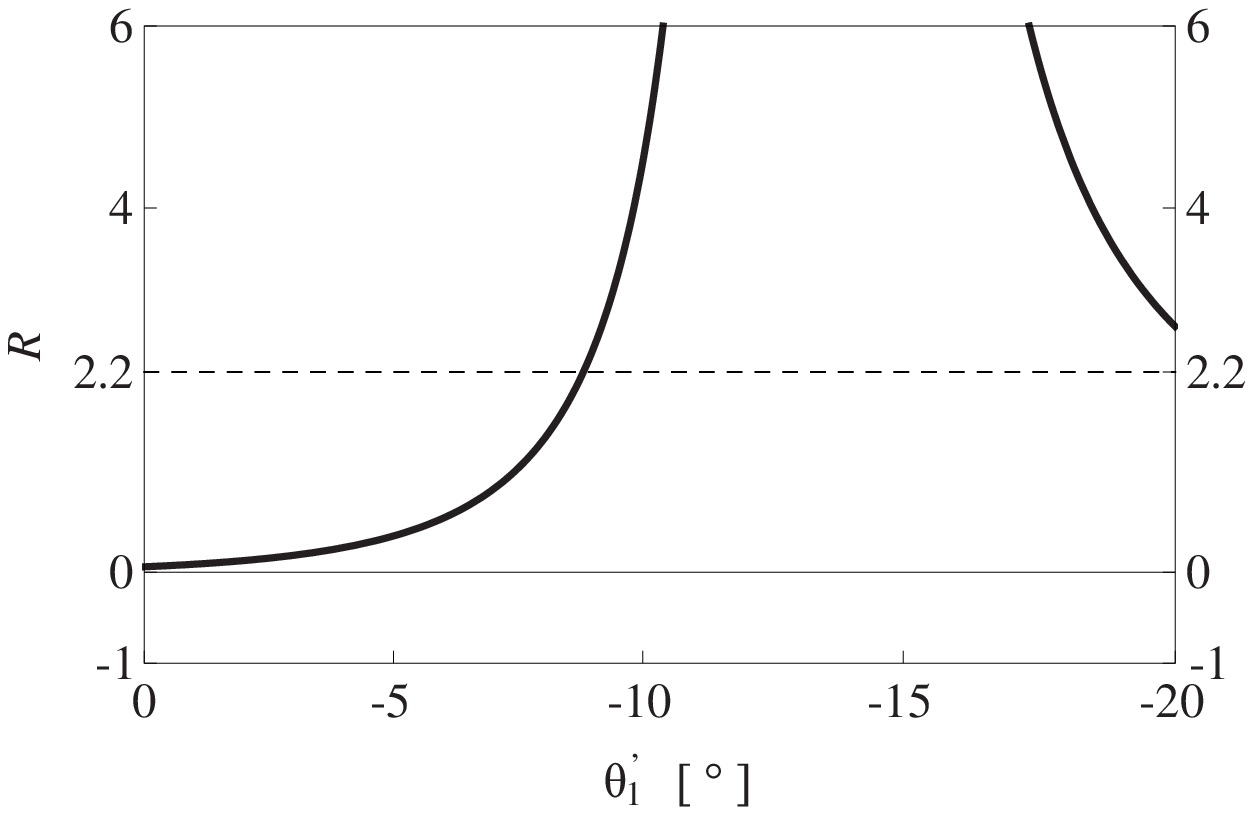}}}
\end{center}
\caption{The ratios
(a) $\mathcal{R}_{\psi\pi} \equiv {\Gamma(|x_{qc}1_{\bar q \bar c} ; 1^{+-} \rangle \rightarrow J/\psi\pi) \over \Gamma(|0_{qc}1_{\bar q \bar c}; 1^{+-} \rangle \rightarrow J/\psi\pi)}$,
(b) $\mathcal{R}_{\eta_c\rho} \equiv {\Gamma(|x_{qc}1_{\bar q \bar c} ; 1^{+-} \rangle \rightarrow \eta_c\rho) \over \Gamma(|0_{qc}1_{\bar q \bar c}; 1^{+-} \rangle \rightarrow \eta_c\rho)}$,
and
(c) $\mathcal{R} \equiv {\mathcal{B}(|x_{qc}1_{\bar q \bar c} ; 1^{+-} \rangle \rightarrow \eta_c\rho) \over \mathcal{B}(|x_{qc}1_{\bar q \bar c} ; 1^{+-} \rangle \rightarrow J/\psi\pi)}$
as functions of the mixing angle $\theta^\prime_1$.}
\label{fig:ratio}
\end{figure*}

We try to add this $|1_{qc}1_{\bar q \bar c} ; 1^{+-} \rangle$ component in this subsection. The interpolating current having the identical internal structure is
\begin{equation}
\eta^{\mathcal Z^\prime}_\mu(x,y) = \eta^3_\mu([uc][\bar d \bar c]) - \eta^4_\mu([uc][\bar d \bar c]) \, ,
\end{equation}
so that $|x_{qc}1_{\bar q \bar c} ; 1^{+-} \rangle$ can be described by
\begin{equation}
\eta^{\rm mix}_\mu(x,y) = \cos \theta^\prime_1~\eta^{\mathcal Z}_\mu(x,y) + i \sin \theta^\prime_1~\eta^{\mathcal Z^\prime}_\mu(x,y)\, ,
\label{eq:diquarkZmix}
\end{equation}
which transforms according to Eq.~(\ref{eq:fierz1}) as:
\begin{eqnarray}
&& \eta^{\rm mix}_\mu(x,y)
\\ \nonumber &\Longrightarrow& + \left( -{i\over3} \cos \theta^\prime_1 + i \sin \theta^\prime_1 \right)  ~ I^{P}(x^\prime)     ~ J^{V}_\mu(y^\prime)
\\ \nonumber && + \left( + {i\over3} \cos \theta^\prime_1 + i \sin \theta^\prime_1 \right)                ~ I^{V}_\mu(x^\prime) ~ J^{P}(y^\prime)
\\ \nonumber && + \left( + {i\over3} \cos \theta^\prime_1 - {i\over3} \sin \theta^\prime_1 \right)        ~ I^{A,\nu}(x^\prime) ~ J^{T}_{\mu\nu}(y^\prime)
\\ \nonumber && + \left( - {i\over3} \cos \theta^\prime_1 - {i\over3} \sin \theta^\prime_1 \right)        ~ I^{T}_{\mu\nu}(x^\prime) ~ J^{A,\nu}(y^\prime) + \cdots \, .
\end{eqnarray}
Note that the two mixing angles $\theta_1$ and $\theta^\prime_1$ are not necessarily the same (probably not the same), but they can be related to each other, {\it i.e.},
\begin{equation}
\theta_1 = f(\theta^\prime_1) \, .
\end{equation}
To solve this relation, we need to know the couplings of $\eta^{\mathcal Z}_\mu$ and $\eta^{\mathcal Z^\prime}_\mu$ to $|0_{qc}1_{\bar q \bar c} ; 1^{+-} \rangle$ and $|1_{qc}1_{\bar q \bar c} ; 1^{+-} \rangle$, which we shall not investigate in the present study.
Anyway, we can plot the three ratios:
\begin{eqnarray}
\nonumber \mathcal{R}_{\psi \pi} &\equiv& {\Gamma(|x_{qc}1_{\bar q \bar c} ; 1^{+-} \rangle \rightarrow J/\psi\pi) \over \Gamma(|0_{qc}1_{\bar q \bar c} ; 1^{+-} \rangle \rightarrow J/\psi\pi)} \, ,
\\ \mathcal{R}_{\eta_c \rho} &\equiv& {\Gamma(|x_{qc}1_{\bar q \bar c} ; 1^{+-} \rangle \rightarrow \eta_c \rho) \over \Gamma(|0_{qc}1_{\bar q \bar c} ; 1^{+-} \rangle \rightarrow \eta_c \rho)} \, ,
\\ \nonumber \mathcal{R} &\equiv& {\mathcal{B}(|x_{qc}1_{\bar q \bar c} ; 1^{+-} \rangle \rightarrow \eta_c\rho) \over \mathcal{B}(|x_{qc}1_{\bar q \bar c} ; 1^{+-} \rangle \rightarrow J/\psi\pi)} \, ,
\end{eqnarray}
as functions of the mixing angle $\theta^\prime_1$, which are shown in Fig.~\ref{fig:ratio}. We find that $\mathcal{R}_{\psi \pi}$ decreases and $\mathcal{R}_{\eta_c \rho}$ increases, so that the ratio $\mathcal{R}$ increases rapidly, as the mixing angle $\theta^\prime_1$ decreasing from $0$ to $-10^{\rm o}$.

Especially, after fine-tuning $\theta^\prime_1 = -8.8^{\rm o}$, we obtain
\begin{eqnarray}
\nonumber \mathcal{R} \equiv {\mathcal{B}(|x_{qc}1_{\bar q \bar c} ; 1^{+-} \rangle \rightarrow \eta_c\rho) \over \mathcal{B}(|x_{qc}1_{\bar q \bar c} ; 1^{+-} \rangle \rightarrow J/\psi\pi)} &=& 2.2 \, ,
\\ {\mathcal{B}(|x_{qc}1_{\bar q \bar c} ; 1^{+-} \rangle \rightarrow h_c\pi) \over \mathcal{B}(|x_{qc}1_{\bar q \bar c} ; 1^{+-} \rangle \rightarrow J/\psi\pi)} &=& 0.052 \, ,
\\ \nonumber {\mathcal{B}(|x_{qc}1_{\bar q \bar c} ; 1^{+-} \rangle \rightarrow \chi_{c1}\rho \rightarrow \chi_{c1} \pi \pi) \over \mathcal{B}(|x_{qc}1_{\bar q \bar c} ; 1^{+-} \rangle \rightarrow J/\psi\pi)} &=& 1.5 \times 10^{-5} \, .
\end{eqnarray}
The first ratio $\mathcal{R}$ is 2.2, which is the same as the BESIII measurement ${\mathcal R}_{Z_c} = 2.2 \pm 0.9$~\cite{Ablikim:2019ipd}.

The decay of $|x_{qc}1_{\bar q \bar c} ; 1^{+-} \rangle$ into two charmed mesons can be described by the current $\eta^{\rm mix}_\mu(x,y)$ together with the transformation~(\ref{eq:fierz2}):
\begin{eqnarray}
&& \eta^{\rm mix}_\mu(x,y)
\\ \nonumber &\Longrightarrow& - {i\over3}~\cos \theta^\prime_1~\xi_\mu^2(x^\prime,y^\prime) + {1\over3}~\cos \theta^\prime_1~\xi_\mu^3(x^\prime,y^\prime)
\\ \nonumber && -~\sin \theta^\prime_1~\xi_\mu^1(x^\prime,y^\prime) - {i\over3}~\sin \theta^\prime_1~\xi_\mu^4(x^\prime,y^\prime)  + \cdots \, ,
\end{eqnarray}
so that
\begin{eqnarray}
\nonumber && {\mathcal{B}(|x_{qc}1_{\bar q \bar c} ; 1^{+-} \rangle \rightarrow D \bar D^{*} + \bar D D^{*}) \over \mathcal{B}(|x_{qc}1_{\bar q \bar c} ; 1^{+-} \rangle \rightarrow J/\psi\pi + \eta_c\rho)} = 0.26 \times {c_2^2 \over c_1^2} \, ,
\\ \nonumber && {\mathcal{B}(|x_{qc}1_{\bar q \bar c} ; 1^{+-} \rangle \rightarrow D \bar D_0^{*} + \bar D D_0^{*} \rightarrow D \bar D \pi) \over \mathcal{B}(|x_{qc}1_{\bar q \bar c} ; 1^{+-} \rangle \rightarrow J/\psi\pi + \eta_c\rho)}
\\ &=& 2.5 \times 10^{-7} \times {c_2^2 \over c_1^2} \, .
\end{eqnarray}
Hence, $|x_{qc}1_{\bar q \bar c} ; 1^{+-} \rangle$ can decay into the $D \bar D^*$ final state, which is consistent with the BESIII observations~\cite{Ablikim:2013xfr,Ablikim:2015swa}. Moreover, it was proposed in Ref.~\cite{Maiani:2017kyi} that: to enable the decay of the $Z_c(3900)$, a constituent of a diquark must tunnel through the barrier of the diquark-antidiquark potential, but this tunnelling for heavy quarks is exponentially suppressed compared to that for light quarks, so the compact tetraquark couplings are expected to favour the open charm modes with respect to charmonium ones. According to this, $c_2$ may be significantly larger than $c_1$, so that $|x_{qc}1_{\bar q \bar c} ; 1^{+-} \rangle$ may mainly decay into two charmed mesons.

\section{Decay properties of the $Z_c(3900)$ as a hadronic molecular state}
\label{sec:decaymolecule}

Another possible interpretation of the $Z_c(3900)$ is the $D \bar D^*$ hadronic molecular state of $J^{PC} = 1^{+-}$~\cite{Voloshin:1976ap,Voloshin:2013dpa,Wang:2013cya,Guo:2013sya}, {\it i.e.}, $| D \bar D^*; 1^{+-} \rangle$ defined in Eq.~(\ref{eq:moleculeZ}). Its relevant current $\xi^{\mathcal Z}_\mu(x,y)$ has been given in Eq.~(\ref{eq:molecule}). We can transform this current to $\theta_\mu^i(x,y)$ according to the transformation~(\ref{eq:fierz4}), through which we shall extract some decay properties of the $Z_c(3900)$ as a hadronic molecular state in the following subsections.

\subsection{$\xi^{\mathcal Z}_\mu\big([\bar c u][\bar d c]\big) \longrightarrow \theta_\mu^{i}\big([\bar c c] + [\bar d u]\big)$}
\label{sec:molecule1}

%
\begin{figure*}[hbt]
\begin{center}
\includegraphics[width=0.8\textwidth]{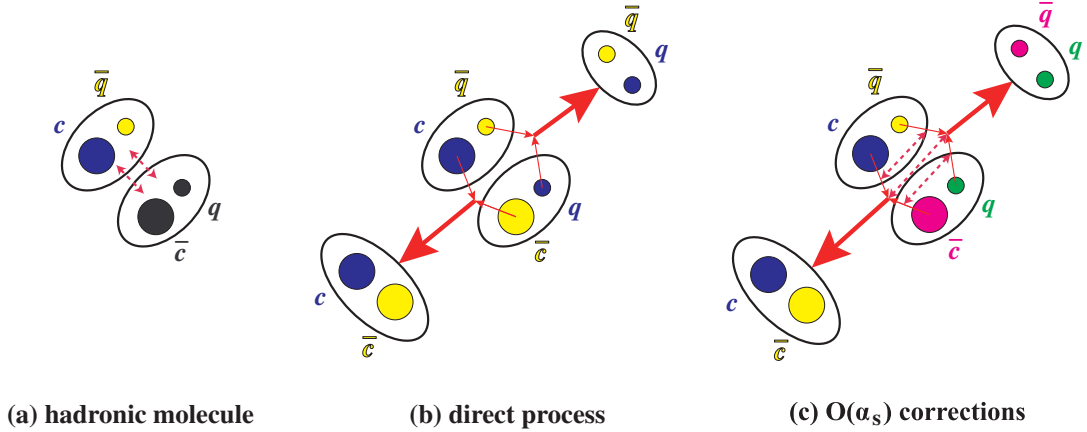}
\caption{The decay of a hadronic molecular state into one charmonium meson and one light meson. This decay can happen through either (b) a direct fall-apart process, or (c) a process with gluon(s) exchanged, that is the $\mathcal{O}(\alpha_s)$ corrections.}
\label{fig:c}
\end{center}
\end{figure*}
%

As depicted in Fig.~\ref{fig:c}, when the $c$ and $\bar c$ quarks meet each other and the $u$ and $\bar d$ quarks meet each other at the same time, a hadronic molecular state can decay into one charmonium meson and one light meson. This process for $| D \bar D^*; 1^{+-} \rangle$ can be described by the transformation~(\ref{eq:fierz4}):
\begin{eqnarray}
&& \xi^{\mathcal Z}_\mu(x,y)
\label{tran:molecule1}
\\ \nonumber &\Longrightarrow& -{1\over6}~\theta_\mu^1(x^\prime,y^\prime) - {1\over6}~\theta_\mu^2(x^\prime,y^\prime)
\\ \nonumber && - {i\over6}~\theta_\mu^3(x^\prime,y^\prime) - {i\over6}~\theta_\mu^4(x^\prime,y^\prime) + \cdots
\\ \nonumber &=& + {i\over6}~I^{P}(x^\prime) ~ J^{V}_\mu(y^\prime) + {i\over6}~I^{V}_\mu(x^\prime) ~ J^{P}(y^\prime)
\\ \nonumber &&  - {i\over6}~I^{A,\nu}(x^\prime) ~ J^{T}_{\mu\nu}(y^\prime) - {i\over6}~I^{T}_{\mu\nu}(x^\prime) ~ J^{A,\nu}(y^\prime) + \cdots \, ,
\end{eqnarray}
where we have only kept the direct fall-apart process described by $\theta_\mu^{1,2,3,4}$, but neglected the $\mathcal{O}(\alpha_s)$ corrections described by $\theta_\mu^{5,6,7,8}$.

We repeat the same procedures as those done in Sec.~\ref{sec:diquark1}, and extract the following coupling constants from this transformation:
\begin{eqnarray}
\nonumber h^S_{\eta_c \rho} &=& {i c_4 \over 6} \lambda_{\eta_c} m_\rho f_{\rho^+} = i c_4~3.65 \times 10^{10}~{\rm MeV}^4 \, ,
\\ \nonumber h^D_{\eta_c \rho} &=& {i c_4\over 6} f_{\eta_c} f^T_{\rho} = i c_4~1.03 \times 10^{4}~{\rm MeV}^2 \, ,
\\ \nonumber h^S_{\psi \pi} &=& {i c_4 \over 6} \lambda_{\pi} m_{J/\psi} f_{J/\psi} = i c_4~5.93 \times 10^{10}~{\rm MeV}^4 \, ,
\\ \nonumber h^D_{\psi \pi} &=& {i c_4 \over 6} f_{\pi^+} f^T_{J/\psi} = i c_4~0.89 \times 10^{4}~{\rm MeV}^2 \, ,
\\ h_{\eta_c b_1} &=& {i c_4 \over 6} f_{\eta_c} f^T_{b_1} = i c_4~1.16 \times 10^{4}~{\rm MeV}^2 \, ,
\\ \nonumber h_{\chi_{c1} \rho} &=& {c_4\over 6} m_{\chi_{c1}} f_{\chi_{c1}} f^T_{\rho} = c_4~3.12 \times 10^{7}~{\rm MeV}^3 \, ,
\\ \nonumber h_{\chi_{c1} b_1} &=& {c_4\over 6} m_{\chi_{c1}} f_{\chi_{c1}} f^T_{b_1} = c_4~3.53 \times 10^{7}~{\rm MeV}^3 \, ,
\\ \nonumber h_{h_c \pi} &=& {i c_4\over 6} f_{\pi^+} f^T_{h_c} = i c_4~0.51 \times 10^{4}~{\rm MeV}^2 \, ,
\\ \nonumber h_{\psi a_1} &=& {c_4\over 6} f^T_{J/\psi} m_{a_1} f_{a_1} = c_4~2.13 \times 10^{7}~{\rm MeV}^3 \, ,
\\ \nonumber h_{h_c a_1} &=& {c_4\over 6} f^T_{h_c} m_{a_1} f_{a_1} = c_4~1.22 \times 10^{7}~{\rm MeV}^3 \, .
\end{eqnarray}
The above coupling constants are related to the $S$- and $D$-wave $| D \bar D^*; 1^{+-} \rangle \to \eta_c \rho$ decays, the $S$- and $D$-wave $| D \bar D^*; 1^{+-} \rangle \rightarrow J/\psi \pi$ decays, and the $| D \bar D^*; 1^{+-} \rangle \rightarrow \eta_c b_1$, $\chi_{c1} \rho$, $\chi_{c1} b_1$, $h_c \pi$, $J/\psi a_1$, $h_c a_1$ decays, respectively. All of them contain an overall factor $c_4$.

Using the above coupling constants, we further obtain
\begin{eqnarray}
\nonumber {\mathcal{B}(| D \bar D^*; 1^{+-} \rangle \rightarrow \eta_c\rho) \over \mathcal{B}(| D \bar D^*; 1^{+-} \rangle \rightarrow J/\psi\pi)} &=& 0.059 \, ,
\label{eq:diquarkbrmixing}
\\ {\mathcal{B}(| D \bar D^*; 1^{+-} \rangle \rightarrow h_c\pi) \over \mathcal{B}(| D \bar D^*; 1^{+-} \rangle \rightarrow J/\psi\pi)} &=& 0.0088 \, ,
\\ \nonumber {\mathcal{B}(| D \bar D^*; 1^{+-} \rangle \rightarrow \chi_{c1}\rho \rightarrow \chi_{c1}\pi \pi) \over \mathcal{B}(| D \bar D^*; 1^{+-} \rangle \rightarrow J/\psi\pi)} &=& 1.4 \times 10^{-6} \, .
\end{eqnarray}
These values are surprisingly the same as Eqs.~(\ref{eq:diquarkbr1}), obtained in Sec.~\ref{sec:diquark1} for the compact tetraquark state $|0_{qc}1_{\bar q \bar c} ; 1^{+-} \rangle$.

\subsection{$\xi^{\mathcal Z}_\mu\big([\bar c u][\bar d c]\big) \longrightarrow \xi^i_\mu\big([\bar c u] + [\bar d c]\big)$}
\label{sec:molecule2}

Assuming the $Z_c(3900)$ to be the $D \bar D^*$ hadronic molecular state of $J^{PC} = 1^{+-}$, it can naturally decay to the $D \bar D^*$ final state, which fall-apart process can be described by itself:
\begin{equation}
\xi^{\mathcal Z}_\mu(x,y) \Longrightarrow \xi_\mu^1(x^\prime,y^\prime) = -i~O^{V}_\mu(x^\prime) ~ O^{P}(y^\prime) + \{ \gamma_\mu \leftrightarrow \gamma_5 \} \, .
\label{tran:molecule2}
\end{equation}
The decay of $| D \bar D^*; 1^{+-} \rangle$ into the $D \bar D^{*}$ final state is contributed by this term to be
\begin{eqnarray}
\nonumber \langle Z_c^+(p,\epsilon) | D^+(p_1) \bar D^{*0}(p_2,\epsilon_2) \rangle &\approx& -ic_5~\lambda_D m_{D^*} f_{D^*}~\epsilon \cdot \epsilon_2
\\ &\equiv& h_{D \bar D^*}~\epsilon \cdot \epsilon_2 \, ,
\\ \nonumber \langle Z_c^+(p,\epsilon) | \bar D^0(p_1) D^{*+}(p_2,\epsilon_2) \rangle &\approx& -ic_5~\lambda_D m_{D^*} f_{D^*}~\epsilon \cdot \epsilon_2
\\ &\equiv& h_{D \bar D^*}~\epsilon \cdot \epsilon_2 \, ,
\end{eqnarray}
where $c_5$ is an overall factor, and it is probably larger than $c_4$. 

Numerically, we obtain
\begin{equation}
h_{D \bar D^*} = -ic_5~2.95 \times 10^{11}~{\rm MeV}^4 \, .
\end{equation}
Comparing the $| D \bar D^*; 1^{+-} \rangle \rightarrow D \bar D^{*}$ decay studied in the present subsection with the $| D \bar D^*; 1^{+-} \rangle \rightarrow J/\psi \pi$ and $| D \bar D^*; 1^{+-} \rangle \rightarrow \eta_c\rho$ decays  studied in the previous subsection, we obtain
\begin{equation}
{\mathcal{B}(| D \bar D^*; 1^{+-} \rangle \rightarrow D \bar D^{*} + \bar D D^{*}) \over \mathcal{B}(| D \bar D^*; 1^{+-} \rangle \rightarrow J/\psi\pi + \eta_c\rho)} = 25 \times {c_5^2 \over c_4^2} \, .
\label{eq:moleculeDD1}
\end{equation}
The current $\xi^{\mathcal Z}_\mu(x,y)$ does not correlate with the term $\xi_\mu^{3} = O^{A}_\mu \times O^{S}$, so that $| D \bar D^*; 1^{+-} \rangle$ does not decay into the $D \bar D_0^{*}$ final state:
\begin{equation}
{\mathcal{B}(| D \bar D^*; 1^{+-} \rangle \rightarrow D \bar D_0^{*} + \bar D D_0^{*} \rightarrow D \bar D \pi) \over \mathcal{B}(| D \bar D^*; 1^{+-} \rangle \rightarrow J/\psi\pi + \eta_c\rho)} \approx 0 \, .
\label{eq:moleculeDD2}
\end{equation}
Eqs.~(\ref{eq:moleculeDD1}) and (\ref{eq:moleculeDD2}) suggest that $| D \bar D^*; 1^{+-} \rangle$ mainly decays into two charmed mesons, other than one charmonium meson and one light meson. This conclusion is opposite to the one obtained in Sec.~\ref{sec:diquark2} for the compact tetraquark state $|0_{qc}1_{\bar q \bar c} ; 1^{+-} \rangle$.

\subsection{Mixing with the $| D^* \bar D^*; 1^{+-} \rangle$}
\label{sec:moleculemixing}

Similarly to Sec.~\ref{sec:diquarkmixing}, we add a small $| D^* \bar D^*; 1^{+-} \rangle$ component
\begin{equation}
| D^* \bar D^*; 1^{+-} \rangle = | D^* \bar D^* \rangle_{J=1} \, ,
\end{equation}
to $| D \bar D^*; 1^{+-} \rangle$ in this subsection to reevaluate the ratio ${\mathcal R}_{Z_c}$. The interpolating current having the same internal structure as $| D^* \bar D^*; 1^{+-} \rangle$ is
\begin{equation}
\xi^{\mathcal Z^\prime}_\mu(x,y) = \xi^2_\mu([\bar c u][\bar d c]) \, ,
\label{eq:moleculeZp}
\end{equation}
so that we can use
\begin{equation}
\xi^{\rm mix}_\mu(x,y) = \cos \theta^\prime_2~\xi^{\mathcal Z}_\mu(x,y) + i \sin \theta^\prime_2~ \xi^{\mathcal Z^\prime}_\mu(x,y) \, ,
\label{eq:moleculemix}
\end{equation}
to described the mixed molecular state
\begin{equation}
|D^{(*)} \bar D^*; 1^{+-} \rangle = \cos \theta_2 ~ |D \bar D^*; 1^{+-} \rangle + \sin \theta_2 ~ |D^* \bar D^*; 1^{+-} \rangle \, .
\label{eq:moleculemixing}
\end{equation}

The current $\xi^{\rm mix}_\mu(x,y)$ transforms according to Eq.~(\ref{eq:fierz4}) to be:
\begin{eqnarray}
&& \xi^{\rm mix}_\mu(x,y)
\\ \nonumber &\Longrightarrow& + \left( + {i\over6}\cos \theta^\prime_2 - {i\over2}\sin \theta^\prime_2 \right) ~ I^{P}(x^\prime)           ~ J^{V}_\mu(y^\prime)
\\ \nonumber && + \left( + {i\over6}\cos \theta^\prime_2 + {i\over2} \sin \theta^\prime_2 \right)               ~ I^{V}_\mu(x^\prime)       ~ J^{P}(y^\prime)
\\ \nonumber && + \left( - {i\over6}\cos \theta^\prime_2 + {i\over6} \sin \theta^\prime_2 \right)               ~ I^{A,\nu}(x^\prime)       ~ J^{T}_{\mu\nu}(y^\prime)
\\ \nonumber && + \left( - {i\over6}\cos \theta^\prime_2 - {i\over6} \sin \theta^\prime_2 \right)               ~ I^{T}_{\mu\nu}(x^\prime)  ~ J^{A,\nu}(y^\prime) + \cdots \, .
\end{eqnarray}
After fine-tuning $\theta^\prime_2 = -8.8^{\rm o}$, we obtain
\begin{eqnarray}
\nonumber {\mathcal R}^\prime \equiv {\mathcal{B}(|D^{(*)} \bar D^*; 1^{+-} \rangle \rightarrow \eta_c\rho) \over \mathcal{B}(|D^{(*)} \bar D^*; 1^{+-} \rangle \rightarrow J/\psi\pi)} &=& 2.2 \, ,
\label{eq:moleculebrmixing}
\\ {\mathcal{B}(|D^{(*)} \bar D^*; 1^{+-} \rangle \rightarrow h_c\pi) \over \mathcal{B}(|D^{(*)} \bar D^*; 1^{+-} \rangle \rightarrow J/\psi\pi)} &=& 0.052 \, ,
\\ \nonumber {\mathcal{B}(|D^{(*)} \bar D^*; 1^{+-} \rangle \rightarrow \chi_{c1}\rho \rightarrow \chi_{c1} \pi \pi) \over \mathcal{B}(|D^{(*)} \bar D^*; 1^{+-} \rangle \rightarrow J/\psi\pi)} &=& 1.5 \times 10^{-5} \, ,
\end{eqnarray}
which values are the same as Eqs.~(\ref{eq:diquarkbrmixing}), obtained in Sec.~\ref{sec:diquark1} for the mixed compact tetraquark state $|x_{qc}1_{\bar q \bar c} ; 1^{+-} \rangle$. Actually, we can also plot the following three ratios
\begin{eqnarray}
\nonumber \mathcal{R}^\prime_{\psi \pi} &\equiv& {\Gamma(|D^{(*)} \bar D^*; 1^{+-} \rangle \rightarrow J/\psi\pi) \over \Gamma(|D \bar D^*; 1^{+-} \rangle \rightarrow J/\psi\pi)} \, ,
\\ \mathcal{R}^\prime_{\eta_c \rho} &\equiv& {\Gamma(|D^{(*)} \bar D^*; 1^{+-} \rangle \rightarrow \eta_c \rho) \over \Gamma(|D \bar D^*; 1^{+-} \rangle \rightarrow \eta_c \rho)} \, ,
\\ \nonumber {\mathcal R}^\prime &\equiv& {\mathcal{B}(|D^{(*)} \bar D^*; 1^{+-} \rangle \rightarrow \eta_c \rho) \over \mathcal{B}(|D^{(*)} \bar D^*; 1^{+-} \rangle \rightarrow J/\psi\pi)} \, ,
\end{eqnarray}
as functions of the mixing angle $\theta^\prime_2$, and the obtained figures are just identical to Fig.~\ref{fig:ratio}, where $\mathcal{R}_{\psi \pi}$, $\mathcal{R}_{\eta_c \rho}$, and ${\mathcal R}$ are shown as functions of $\theta^\prime_1$.

We also obtain
\begin{eqnarray}
&& {\mathcal{B}(|D^{(*)} \bar D^*; 1^{+-} \rangle \rightarrow D \bar D^{*} + \bar D D^{*}) \over \mathcal{B}(|D^{(*)} \bar D^*; 1^{+-} \rangle \rightarrow J/\psi\pi + \eta_c\rho)} = 67 \times {c_5^2 \over c_4^2} \, ,
\\ \nonumber && {\mathcal{B}(|D^{(*)} \bar D^*; 1^{+-} \rangle \rightarrow D \bar D_0^{*} + \bar D D_0^{*} \rightarrow D \bar D \pi) \over \mathcal{B}(|D^{(*)} \bar D^*; 1^{+-} \rangle \rightarrow J/\psi\pi + \eta_c\rho)} \approx 0 \, ,
\end{eqnarray}
suggesting that $|D^{(*)} \bar D^*; 1^{+-} \rangle$ mainly decays into two charmed mesons.

\section{Summary and discussions}
\label{sec:summary}

\begin{table*}[hbt]
\begin{center}
\renewcommand{\arraystretch}{1.5}
\caption{Relative branching ratios of the $Z_c(3900)$ evaluated through the Fierz rearrangement. $\theta_{1,2}^\prime$ are the two mixing angles defined in Eqs.~(\ref{eq:diquarkZmix}) and (\ref{eq:moleculemix}), which are fine-tuned to be $\theta^\prime_1 = \theta^\prime_2 = -8.8^{\rm o}$, so that ${\mathcal{B}(|x_{qc}1_{\bar q \bar c} ; 1^{+-} \rangle \rightarrow \eta_c\rho) \over \mathcal{B}(|x_{qc}1_{\bar q \bar c} ; 1^{+-} \rangle \rightarrow J/\psi\pi)} = {\mathcal{B}(|D^{(*)} \bar D^*; 1^{+-} \rangle \rightarrow \eta_c\rho) \over \mathcal{B}(|D^{(*)} \bar D^*; 1^{+-} \rangle \rightarrow J/\psi\pi)} = 2.2$~\cite{Ablikim:2019ipd}. In this table we do not take into account the phase angle $\phi$ between $S$- and $D$-wave coupling constants. }
\begin{tabular}{ c | c | c | c | c}
\hline\hline
\multirow{2}{*}{~~~~~~~~~~~Channels~~~~~~~~~~~} & \multirow{2}{*}{~~~~~~$|0_{qc}1_{\bar q \bar c}; 1^{+-} \rangle$~~~~~~} & ~~~~~$|x_{qc}1_{\bar q \bar c} ; 1^{+-} \rangle$~~~~~ & \multirow{2}{*}{~~~$|D \bar D^*; 1^{+-} \rangle$~~~} & ~~~$|D^{(*)} \bar D^*; 1^{+-} \rangle$~~~
\\ & &  ($\theta^\prime_1 = -8.8^{\rm o}$) & & ($\theta^\prime_2 = -8.8^{\rm o}$)
\\ \hline\hline
${\mathcal{B}(Z_c \rightarrow \eta_c\rho) \over \mathcal{B}(Z_c \rightarrow J/\psi\pi)}$
& $0.059$ & $2.2$~(input) & $0.059$ & $2.2$~(input)
\\ \hline
${\mathcal{B}(Z_c \rightarrow h_c\pi) \over \mathcal{B}(Z_c \rightarrow J/\psi\pi)}$
& $0.0088$ &  $0.052$  & $0.0088$ &  $0.052$
\\ \hline
${\mathcal{B}(Z_c \rightarrow \chi_{c1}\rho \rightarrow \chi_{c1} \pi \pi) \over \mathcal{B}(Z_c \rightarrow J/\psi\pi)}$
& $1.4 \times 10^{-6}$ &  $1.5 \times 10^{-5}$  & $1.4 \times 10^{-6}$ &  $1.5 \times 10^{-5}$
\\ \hline \hline
${\mathcal{B}(Z_c \rightarrow D \bar D^{*} + \bar D D^{*}) \over \mathcal{B}(Z_c \rightarrow J/\psi\pi + \eta_c\rho)}$
& $\approx 0$ &  $0.26~t_1$ & $25~t_2$ & $67~t_2$
\\ \hline
${\mathcal{B}(Z_c \rightarrow D \bar D_0^{*} + \bar D D_0^{*} \rightarrow D \bar D \pi) \over \mathcal{B}(Z_c \rightarrow J/\psi\pi + \eta_c\rho)}$
& $9.3~t_1 \times 10^{-8}$ &  $2.5~t_1 \times 10^{-7}$   & $\approx 0$ & $\approx 0$
\\ \hline\hline
\end{tabular}
\label{tab:relative}
\end{center}
\end{table*}

In this paper we systematically construct all the tetraquark currents/operators of $J^{PC} = 1^{+-}$ with the quark content $c \bar c q \bar q$ ($q=u/d$). There are three configurations: $[cq][\bar c \bar q]$, $[\bar c q][\bar q c]$, and $[\bar c c][\bar q q]$, and for each configuration we construct eight independent currents. We use the Fierz rearrangement of the Dirac and color indices to derive their relations, through which we study strong decay properties of the $Z_c(3900)$:
\begin{itemize}

\item Using the transformation of $[qc][\bar q \bar c] \to [\bar c c][\bar q q]$, we study decay properties of the $Z_c(3900)$ as a compact diquark-antidiquark tetraquark state into one charmonium meson and one light meson.

\item Using the transformation of $[qc][\bar q \bar c] \to [\bar c q][\bar q c]$, we study decay properties of the $Z_c(3900)$ as a compact diquark-antidiquark tetraquark state into two charmed mesons.

\item We use the transformation of the $[qc][\bar q \bar c]$ currents to the color-singlet-color-singlet $[\bar c c][\bar q q]$ and $[\bar c q][\bar q c]$ currents, and obtain the same relative branching ratios as the above results.

\item Using the transformation of $[\bar c q][\bar q c] \to [\bar c c][\bar q q]$, we study decay properties of the $Z_c(3900)$ as a hadronic molecular state into one charmonium meson and one light meson.

\item Through the $[\bar c q][\bar q c]$ currents themselves, we study decay properties of the $Z_c(3900)$ as a hadronic molecular state into two charmed mesons.

\end{itemize}

Our results suggest that the possible decay channels of the $Z_c(3900)$ are: a) the two-body decays $Z_c(3900) \to J/\psi\pi$, $Z_c(3900) \to \eta_c\rho$, $Z_c(3900) \to h_c\pi$, and $Z_c(3900) \to D \bar D^{*}$, b) the three-body decays $Z_c(3900) \rightarrow \chi_{c1}\rho \rightarrow \chi_{c1}\pi \pi$ and $Z_c(3900) \rightarrow D \bar D_0^{*} + \bar D D_0^{*} \rightarrow D \bar D \pi$, and c) the many-body decay chains $Z_c(3900) \rightarrow J/\psi a_1 \rightarrow J/\psi \rho \pi \rightarrow J/\psi + 3 \pi$ and $Z_c(3900) \rightarrow \eta_c b_1 \rightarrow \eta_c \omega \pi \rightarrow \eta_c + 4 \pi$. Their relative branching ratios are summarized in Table~\ref{tab:relative}, where we have investigated the following interpretations of the $Z_c(3900)$:
\begin{widetext}
\begin{itemize}

\item In the second and third columns of Table~\ref{tab:relative}, $|0_{qc}1_{\bar q \bar c} ; 1^{+-} \rangle$ and $|x_{qc}1_{\bar q \bar c} ; 1^{+-} \rangle$ denote the compact tetraquark states of $J^{PC} = 1^{+-}$, defined in Eq.~(\ref{eq:diquarkZ}) and Eq.~(\ref{eq:diquarkmixing}), respectively. Especially, we have considered the mixing between the compact tetraquarks states
\begin{equation}
| 0_{qc}1_{\bar q \bar c}; 1^{+-} \rangle \oplus | 1_{qc}1_{\bar q \bar c}; 1^{+-} \rangle \rightarrow | x_{qc}1_{\bar q \bar c}; 1^{+-} \rangle \, .
\label{eq:mix1}
\end{equation}
Using the mixing angle $\theta^\prime_1 = -8.8^{\rm o}$, we obtain
\begin{eqnarray}
\nonumber && {\mathcal{B}\left(| x_{qc}1_{\bar q \bar c}; 1^{+-} \rangle \rightarrow ~J/\psi\pi~
:~~~~~~ \eta_c\rho ~~~~~~
:~~ h_c\pi ~~\,
:~ \chi_{c1}\rho (\rightarrow \pi \pi)~
:~~~ D \bar D^{*}~~~
:~ D \bar D_0^{*} (\rightarrow \bar D \pi)~
\right) \over \mathcal{B}(| x_{qc}1_{\bar q \bar c}; 1^{+-} \rangle \rightarrow J/\psi\pi)}
\\ \label{eq:result1} &\approx&
~~~~~~~~~~~~~~~~~~~~~~~~~~~~~~1~~~\, : ~2.2~({\rm input})~ : ~~0.05~~ : ~~~~~~10^{-5}~~~~~\, : ~~0.82~t_1~~ : ~~~~~10^{-6}~t_1~\, .
\end{eqnarray}

\item In the fourth and fifth columns of Table~\ref{tab:relative}, $| D \bar D^*; 1^{+-} \rangle$ and $|D^{(*)} \bar D^*; 1^{+-} \rangle$ denote the hadronic molecular states of $J^{PC} = 1^{+-}$, defined in Eq.~(\ref{eq:moleculeZ}) and  Eq.~(\ref{eq:moleculemixing}), respectively. Especially, we have considered the mixing between the hadronic molecule states
\begin{equation}
| D \bar D^*; 1^{+-} \rangle \oplus |D^{*} \bar D^*; 1^{+-} \rangle \rightarrow  |D^{(*)} \bar D^*; 1^{+-} \rangle \, .
\label{eq:mix2}
\end{equation}
Using the mixing angle $\theta^\prime_2 = -8.8^{\rm o}$, we obtain
\begin{eqnarray}
\nonumber && {\mathcal{B}\left(|D^{(*)} \bar D^*; 1^{+-} \rangle \rightarrow ~J/\psi\pi~
:~~~~~~ \eta_c\rho ~~~~~~
:~~ h_c\pi ~~\,
:~ \chi_{c1}\rho (\rightarrow \pi \pi)~
:~~~ D \bar D^{*}~
\right) \over \mathcal{B}(|D^{(*)} \bar D^*; 1^{+-} \rangle \rightarrow J/\psi\pi)}
\\ \label{eq:result2} &\approx&
~~~~~~~~~~~~~~~~~~~~~~~~~~~~~~~\,1~~~\, : ~2.2~({\rm input})~ : ~~0.05~~ : ~~~~~~10^{-5}~~~~~\, : ~~~210~t_2~\, .
\end{eqnarray}

\end{itemize}
\end{widetext}
In the above expressions, we have used the recent BESIII measurement ${\mathcal R}_{Z_c} \equiv {\mathcal{B}(Z_c(3900) \rightarrow \eta_c\rho) \over \mathcal{B}(Z_c(3900) \rightarrow J/\psi\pi)} = 2.2 \pm 0.9$~\cite{Ablikim:2019ipd} as an input to determine the mixing angles $\theta^\prime_1$ and $\theta^\prime_2$. The ratio $t_1 \equiv {c_2^2 / c_1^2}$ is the parameter measuring which process happens more easily, the process depicted in Fig.~\ref{fig:a}(b) or the process depicted in Fig.~\ref{fig:b}(b). Generally speaking, the exchange of one light quark with another light quark seems to be easier than the exchange of one light quark with another heavy quark~\cite{Landau,Maiani:2017kyi}, so it can be the case that $t_1 \geq 1$. As discussed in Sec.~\ref{sec:molecule2}, $c_5$ is probably larger than $c_4$, so that the other ratio $t_2 \equiv {c_5^2 / c_4^2} \geq 1$.

The above relative branching ratios calculated in the present study turn out to be very much different, which might be one of the reasons why many multiquark states were observed only in a few decay channels~\cite{Chen:2016qju}. Note that in order to extract the above results, we have only considered the leading-order fall-apart decays described by color-singlet-color-singlet meson-meson currents, but neglected the $\mathcal{O}(\alpha_s)$ corrections described by color-octet-color-octet meson-meson currents, so there can be other decay channels.

Based on Table~\ref{tab:relative} as well as Eqs.~(\ref{eq:result1}) and (\ref{eq:result2}), we conclude this paper:
\begin{itemize}

\item The relative branching ratios ${\mathcal{B}(|0_{qc}1_{\bar q \bar c} ; 1^{+-} \rangle \rightarrow \eta_c\rho) \over \mathcal{B}(|0_{qc}1_{\bar q \bar c} ; 1^{+-} \rangle \rightarrow J/\psi\pi)}$ and ${\mathcal{B}(| D \bar D^*; 1^{+-} \rangle \rightarrow \eta_c\rho) \over \mathcal{B}(| D \bar D^*; 1^{+-} \rangle \rightarrow J/\psi\pi)}$ are both around 0.059, significantly smaller than the BESIII measurement ${\mathcal R}_{Z_c} = 2.2 \pm 0.9$ at $\sqrt{s} = 4.226$~GeV~\cite{Ablikim:2019ipd}. However, we can add a small $|1_{qc}1_{\bar q \bar c} ; 1^{+-} \rangle$ component to $|0_{qc}1_{\bar q \bar c} ; 1^{+-} \rangle$ to obtain ${\mathcal{B}(|x_{qc}1_{\bar q \bar c} ; 1^{+-} \rangle \rightarrow \eta_c\rho) \over \mathcal{B}(|x_{qc}1_{\bar q \bar c} ; 1^{+-} \rangle \rightarrow J/\psi\pi)} = 2.2$; we can also add a small $|D^{*} \bar D^*; 1^{+-} \rangle$ component to $| D \bar D^*; 1^{+-} \rangle$ to obtain ${\mathcal{B}(|D^{(*)} \bar D^*; 1^{+-} \rangle \rightarrow \eta_c\rho) \over \mathcal{B}(|D^{(*)} \bar D^*; 1^{+-} \rangle \rightarrow J/\psi\pi)} = 2.2$. Note that if the relevant mixing angles change dynamically, the ratio ${\mathcal R}_{Z_c}$ would also change dynamically.

\item Relative branching ratios of the $| D \bar D^*; 1^{+-} \rangle$ decays into one charmonium meson and one light meson are the same as those of the $|0_{qc}1_{\bar q \bar c} ; 1^{+-} \rangle$ decays. After taking proper mixing angles, relative branching ratios of the $|D^{(*)} \bar D^*; 1^{+-} \rangle$ decays into one charmonium meson and one light meson are also the same as those of the $|x_{qc}1_{\bar q \bar c} ; 1^{+-} \rangle$ decays. This suggests that one may not discriminate between the compact tetraquark and hadronic molecule scenarios by only investigating relative branching ratios of the $Z_c(3900)$ decays into one charmonium meson and one light meson.

\item $|0_{qc}1_{\bar q \bar c} ; 1^{+-} \rangle$ mainly decays into one charmonium meson and one light meson, but $|x_{qc}1_{\bar q \bar c} ; 1^{+-} \rangle$ might mainly decay into two charmed mesons after taking into account the barrier of the diquark-antidiquark potential (see detailed discussions in Ref.~\cite{Maiani:2017kyi} proposing $c_2 \gg c_1$). Both $| D \bar D^*; 1^{+-} \rangle$ and $|D^{(*)} \bar D^*; 1^{+-} \rangle$ mainly decay into two charmed mesons.

\end{itemize}

It is useful to generally discuss about our uncertainty. In the present study we have worked within the naive factorization scheme, so our uncertainty is larger than the well-developed QCD factorization method~\cite{Beneke:1999br,Beneke:2000ry,Beneke:2001ev}, that is at the 5\% level when being applied to study weak and radiative decay properties of conventional (heavy) hadrons. On the other hand, the tetraquark decay constant $f_{Z_c}$ is removed when calculating relative branching ratios. This significantly reduces our uncertainty, because this parameter has not been well determined yet. Hence, we roughly estimate our uncertainty to be at the $X^{+100\%}_{-~50\%}$ level. 

Now let us compare our results with other theoretical calculations. First we compare them with the QCD sum rule results obtained in Refs.~\cite{Dias:2013xfa,Agaev:2016dev}, where the $Z_c(3900)$ is assumed to be a compact diquark-antidiquark tetraquark state. In the present study we find that decays of the $Z_c(3900)$ into $J/\psi\pi$ and $\eta_c\rho$ can happen through both $S$-wave and $D$-wave, and we have calculated these two amplitudes together, as shown in Eqs.~(\ref{decay:etacrho}-\ref{lag:psipiD}); while we can also calculate them individually and obtain
\begin{eqnarray}
&& {\mathcal{B}(|0_{qc}1_{\bar q \bar c} ; 1^{+-} \rangle \rightarrow \eta_c\rho)_{S\mbox{-}\rm wave} \over \mathcal{B}(|0_{qc}1_{\bar q \bar c} ; 1^{+-} \rangle \rightarrow J/\psi\pi)_{S\mbox{-}\rm wave}} = 0.24 \, ,
\\ \nonumber && {\mathcal{B}(|0_{qc}1_{\bar q \bar c} ; 1^{+-} \rangle \rightarrow \eta_c\rho)_{D\mbox{-}\rm wave} \over \mathcal{B}(|0_{qc}1_{\bar q \bar c} ; 1^{+-} \rangle \rightarrow J/\psi\pi)_{D\mbox{-}\rm wave}} = 0.82 \, .
\end{eqnarray}
In the former equation we have only considered the $S$-wave amplitudes, while in the latter only the $D$-wave ones. The QCD sum rule study in Ref.~\cite{Dias:2013xfa} only considers the $S$-wave amplitudes, where they obtained
\begin{eqnarray}
&& \Gamma(Z_c(3900) \rightarrow \eta_c\rho) = 27.5 \pm 8.5~{\rm MeV} \, ,
\\ \nonumber && \Gamma(Z_c(3900) \rightarrow J/\psi\pi) = 29.1 \pm 8.2~{\rm MeV} \, ,
\end{eqnarray}
so that
\begin{eqnarray}
&& {\mathcal{B}(Z_c(3900)  \rightarrow \eta_c\rho)_{S\mbox{-}\rm wave} \over \mathcal{B}(Z_c(3900)  \rightarrow J/\psi\pi)_{S\mbox{-}\rm wave}} = 0.95^{+0.47}_{-0.36} \, .
\end{eqnarray}
The QCD sum rule study in Ref.~\cite{Agaev:2016dev} only considers the $D$-wave amplitudes, where they obtained
\begin{eqnarray}
&& \Gamma(Z_c(3900) \rightarrow \eta_c\rho) = 23.8 \pm 4.9~{\rm MeV} \, ,
\\ \nonumber && \Gamma(Z_c(3900) \rightarrow J/\psi\pi) = 41.9 \pm 9.4~{\rm MeV} \, ,
\end{eqnarray}
so that
\begin{eqnarray}
&& {\mathcal{B}(Z_c(3900)  \rightarrow \eta_c\rho)_{D\mbox{-}\rm wave} \over \mathcal{B}(Z_c(3900)  \rightarrow J/\psi\pi)_{D\mbox{-}\rm wave}} = 0.57^{+0.20}_{-0.16} \, .
\end{eqnarray}
Hence, our results are more or less consistent with the QCD sum rule calculations~\cite{Dias:2013xfa,Agaev:2016dev}. Here we would like to note that the $D$-wave decay amplitudes are important and can not be neglected:
\begin{eqnarray}
&& {\mathcal{B}(|0_{qc}1_{\bar q \bar c} ; 1^{+-} \rangle \rightarrow \eta_c\rho)_{D\mbox{-}\rm wave} \over \mathcal{B}(|0_{qc}1_{\bar q \bar c} ; 1^{+-} \rangle \rightarrow \eta_c\rho)_{S\mbox{-}\rm wave}} = 0.51 \, ,
\\ \nonumber && {\mathcal{B}(|0_{qc}1_{\bar q \bar c} ; 1^{+-} \rangle \rightarrow J/\psi\pi)_{D\mbox{-}\rm wave} \over \mathcal{B}(|0_{qc}1_{\bar q \bar c} ; 1^{+-} \rangle \rightarrow J/\psi\pi)_{S\mbox{-}\rm wave}} = 0.15 \, .
\end{eqnarray}
Actually, there is still one parameter not considered in our calculations, that is the phase angle $\phi$ between $S$- and $D$-wave decay amplitudes. For completeness, we shall investigate its relevant uncertainty in Appendix~\ref{app:phase}.

Then we compare our results with Ref.~\cite{Esposito:2014hsa}, where the authors assumed the $Z_c(3900)$ to be a hadronic molecular state and used the Non-Relativistic Effective Field Theory (a framework based on HQET and NRQCD) to obtain
\begin{eqnarray}
&& {\mathcal{B}(Z_c(3900) \rightarrow \eta_c\rho) \over \mathcal{B}(Z_c(3900) \rightarrow J/\psi\pi)} = 0.046^{+0.025}_{-0.017} \, .
\end{eqnarray}
This value is well consistent with our result:
\begin{eqnarray}
&& {\mathcal{B}(| D \bar D^*; 1^{+-} \rangle \rightarrow \eta_c\rho) \over \mathcal{B}(| D \bar D^*; 1^{+-} \rangle \rightarrow J/\psi\pi)} = 0.059 \, .
\end{eqnarray}

To end this paper, we propose the BESIII, Belle, Belle-II, and LHCb Collaborations to search for those decay channels not observed yet, in order to better understand the nature of the $Z_c(3900)$.

\section*{Acknowledgments}

We thank Fu-Sheng Yu and Qin Chang for helpful discussions.
This project is supported by the National Natural Science Foundation of China under Grants No.~11722540.

\appendix

\section{Uncertainties due to phase angles}
\label{app:phase}

\begin{table*}[hbt]
\begin{center}
\renewcommand{\arraystretch}{1.5}
\caption{Relative branching ratios of the $Z_c(3900)$ evaluated through the Fierz rearrangement. The two mixing angles are fine-tuned to be $\theta^\prime_1 = \theta^\prime_2 = -10.1^{\rm o}$, so that ${\mathcal{B}(|x_{qc}1_{\bar q \bar c} ; 1^{+-} \rangle \rightarrow \eta_c\rho) \over \mathcal{B}(|x_{qc}1_{\bar q \bar c} ; 1^{+-} \rangle \rightarrow J/\psi\pi)} = {\mathcal{B}(|D^{(*)} \bar D^*; 1^{+-} \rangle \rightarrow \eta_c\rho) \over \mathcal{B}(|D^{(*)} \bar D^*; 1^{+-} \rangle \rightarrow J/\psi\pi)} = 2.2$~\cite{Ablikim:2019ipd}. In this table we fix the phase angle $\theta$ between all the $S$- and $D$-wave coupling constants to be $\theta = \pi$.}
\begin{tabular}{ c | c | c | c | c}
\hline\hline
\multirow{2}{*}{~~~~~~~~~~~Channels~~~~~~~~~~~} & \multirow{2}{*}{~~~~~~$|0_{qc}1_{\bar q \bar c}; 1^{+-} \rangle$~~~~~~} & ~~~~~$|x_{qc}1_{\bar q \bar c} ; 1^{+-} \rangle$~~~~~ & \multirow{2}{*}{~~~$|D \bar D^*; 1^{+-} \rangle$~~~} & ~~~$|D^{(*)} \bar D^*; 1^{+-} \rangle$~~~
\\ & &  ($\theta^\prime_1 = -10.1^{\rm o}$) & & ($\theta^\prime_2 = -10.1^{\rm o}$)
\\ \hline\hline
${\mathcal{B}(Z_c \rightarrow \eta_c\rho) \over \mathcal{B}(Z_c \rightarrow J/\psi\pi)}$
& $0.36$ & $2.2$~(input) & $0.36$ & $2.2$~(input)
\\ \hline
${\mathcal{B}(Z_c \rightarrow h_c\pi) \over \mathcal{B}(Z_c \rightarrow J/\psi\pi)}$
& $0.0018$ &  $0.0038$  & $0.0018$ &  $0.0038$
\\ \hline
${\mathcal{B}(Z_c \rightarrow \chi_{c1}\rho \rightarrow \chi_{c1} \pi \pi) \over \mathcal{B}(Z_c \rightarrow J/\psi\pi)}$
& $2.8 \times 10^{-7}$ &  $1.2 \times 10^{-6}$  & $2.8 \times 10^{-7}$ &  $1.2 \times 10^{-6}$
\\ \hline \hline
${\mathcal{B}(Z_c \rightarrow D \bar D^{*} + \bar D D^{*}) \over \mathcal{B}(Z_c \rightarrow J/\psi\pi + \eta_c\rho)}$
& $\approx 0$ &  $0.059~t_1$ & $3.9~t_2$ & $5.2~t_2$
\\ \hline
${\mathcal{B}(Z_c \rightarrow D \bar D_0^{*} + \bar D D_0^{*} \rightarrow D \bar D \pi) \over \mathcal{B}(Z_c \rightarrow J/\psi\pi + \eta_c\rho)}$
& $1.5~t_1 \times 10^{-8}$ &  $2.0~t_1 \times 10^{-8}$   & $\approx 0$ & $\approx 0$
\\ \hline\hline
\end{tabular}
\label{tab:relative2}
\end{center}
\end{table*}

There are two different effective Lagrangians for the $Z_c(3900)$ decay into the $\eta_c \rho$ final state, as given in Eqs.~(\ref{lag:etcrhoS}) and (\ref{lag:etcrhoD}):
\begin{eqnarray}
\mathcal{L}^S_{\eta_c \rho} &=& g^S_{\eta_c \rho}~Z_{c}^{+,\mu}~\eta_c~\rho^{-}_{\mu} + \cdots \, ,
\\ \mathcal{L}^D_{\eta_c \rho} &=& g^D_{\eta_c \rho} \times \left( g^{\mu\sigma}g^{\nu\rho} - g^{\mu\nu}g^{\rho\sigma} \right)
\\ \nonumber && ~~~~~ \times Z_{c,\mu}^{+}~\partial_\rho \eta_c~\partial_\sigma \rho^{-}_{\nu} + \cdots \, .
\end{eqnarray}
There are also two different effective Lagrangians for the $Z_c(3900)$ decay into the $J/\psi \pi$ final state, as given in Eqs.~(\ref{lag:psipiS}) and (\ref{lag:psipiD}):
\begin{eqnarray}
\mathcal{L}^S_{\psi \pi} &=& g^S_{\psi \pi}~Z_{c}^{+,\mu}~\psi_\mu~\pi^- + \cdots \, ,
\\ \mathcal{L}^D_{\psi \pi} &=& g^D_{\psi \pi} \times \left( g^{\mu\rho}g^{\nu\sigma} - g^{\mu\nu}g^{\rho\sigma} \right)
\\ \nonumber && ~~~~~ \times Z_{c,\mu}^{+}~\partial_\rho \psi_\nu~\partial_\sigma \pi^- + \cdots \, .
\end{eqnarray}
There can be a phase angle $\phi$ between $g^S_{\eta_c \rho}$ and $g^D_{\eta_c \rho}$ as well as between $g^S_{\psi \pi}$ and $g^D_{\psi \pi}$. This parameter is unknown and so not fixed, because in QCD sum rules one can only calculate the modular square of the decay constant, such as $|f_{\eta_c}|^2$. This might also be the case for Lattice QCD and light front model, for example, see the different definitions of $f_{\eta_c}$ in Refs.~\cite{Becirevic:2013bsa,Veliev:2010vd}.

We rotate this phase angle between all the $S$- and $D$-wave coupling constants to be $\phi = \pi$, and redo the previous calculations. The results are summarized in Table~\ref{tab:relative2}. Especially, using the mixing angle $\theta^\prime_1 = \theta^\prime_2 = -10.1^{\rm o}$, we obtain
\begin{widetext}
\begin{eqnarray}
\nonumber && {\mathcal{B}\left(| x_{qc}1_{\bar q \bar c}; 1^{+-} \rangle \rightarrow ~~J/\psi\pi~\,
:~~~~~~ \eta_c\rho ~~~~~~
:~~~ h_c\pi ~~~
:~ \chi_{c1}\rho (\rightarrow \pi \pi)~
:~~~D \bar D^{*}~~~
:~ D \bar D_0^{*} (\rightarrow \bar D \pi)~
\right) \over \mathcal{B}(| x_{qc}1_{\bar q \bar c}; 1^{+-} \rangle \rightarrow J/\psi\pi)}
\\ &\approx&
~~~~~~~~~~~~~~~~~~~~~~~~~~~~~~~1~~~~ : ~2.2~({\rm input})~ : ~~0.004~~ : ~~~~~~10^{-6}~~~~~\, : ~~0.19~t_1~~ : ~~~~~10^{-7}~t_1~\, ,
\\[2mm]
\nonumber && {\mathcal{B}\left(|D^{(*)} \bar D^*; 1^{+-} \rangle \rightarrow ~J/\psi\pi~
:~~~~~~ \eta_c\rho ~~~~~~
:~~~ h_c\pi ~~~
:~ \chi_{c1}\rho (\rightarrow \pi \pi)~
:~~~ D \bar D^{*}~
\right) \over \mathcal{B}(|D^{(*)} \bar D^*; 1^{+-} \rangle \rightarrow J/\psi\pi)}
\\ &\approx&
~~~~~~~~~~~~~~~~~~~~~~~~~~~~~~~\,1~~~\, : ~2.2~({\rm input})~ : ~~0.004~~ : ~~~~~~10^{-6}~~~~~\, : ~~~16~t_2~\, .
\end{eqnarray}
\end{widetext}

\end{document}